\documentclass[%
reprint,
amsmath,amssymb,
aps,
pra,
]{revtex4-2}

\usepackage{graphicx}
\usepackage{dcolumn}
\usepackage{bm}

\begin{document}


\title{
Macroscopic Singlet, Triplet, and Colour-Charged States of Coherent Photons
}


\author{Shinichi Saito}
 \email{shinichi.saito.qt@hitachi.com}
\affiliation{Center for Exploratory Research Laboratory, Research \& Development Group, Hitachi, Ltd. Tokyo 185-8601, Japan.}

\date{\today}

\begin{abstract}
A ray of photons, emitted from a laser source, is in a coherent state, where macroscopic number of photons are degenerate in the same quantum state.
The coherent state has degrees of freedom for spin and orbital angular momentum, which allow an arbitrary superposition state among orthogonal states with varying their amplitudes and phases, described by a representation theory of Lie algebra and Lie group.
Here, we experimentally demonstrate that we can construct generators of rotations for the quantum states of coherent photons, simply by combining widely available optical components, such as half- and quarter-wave plates and vortex lenses.
We have found that a superposition state between vortexed and no-vortex states is characterised by the motion of the topological charge upon the rotation in the SU(3) states.
We also realised singlet and triplet states by combining rays of photons with orthogonal polarisation states and vortexed states.
This corresponds to realise an effective SU(4) state and we have confirmed the projection to an SU(2)$\times$SU(2) state upon passing through a polariser.
\end{abstract}

\maketitle

\section{Introduction}

The notion of macroscopic quantum phenomena is applicable only to the limited number of systems, such as superfluid He \cite{Penrose56,Matsubara56,Yang62,Leggett04}, superconductors \cite{Bardeen57,Bogoljubov58,Schrieffer71,Nozieres85}, cold atoms \cite{Ketterle02,Cornell02},  and lasers \cite{Yariv97,Gil16,Goldstein11,Hecht17,Pedrotti07}.
These systems would be useful as platforms to explore why our macroscopic classical world seems to be different from the microscopic world, governed by quantum mechanics  \cite{Caldeira81,Leggett85,Leggett95,Frowis18}.

These classical examples of macroscopic quantum phenomena might be less exotic, however, compared with Schr\"{o}dinger's cat state \cite{Caldeira81,Leggett85,Leggett95,Frowis18}, since macroscopic number of elementary particles simply occupy the same state, due to the nature of Bose-Einstein Condensation (BEC) upon broken symmetry 
\cite{Higgs64,Goldstone62,Nambu59,Abrikosov75,Fetter03,Wen04,Nagaosa99,Altland10}.
This is easily understood for bosons \cite{Ketterle02,Cornell02,Abrikosov75,Fetter03,Wen04,Nagaosa99,Altland10}, since quantum statistics allows bosons to occupy the same energy state, such that the macroscopic number of bosons simply occupy the same state, if the interaction is weak, to expect BEC at low enough temperatures.
Even for degenerate fermions, such as electrons in a metal with weak attractive interaction, Cooper pairs are formed between time-reversal symmetric spin-up and down electrons, which behave like bosons \cite{Nozieres85}, leading to a spontaneous U(1) (the unitary group of 1-dimension) symmetry breaking to exhibit superconducting behaviour below a critical temperature \cite{Bardeen57,Anderson58,Bogoljubov58,Schrieffer71,Nozieres85}, similar to BEC \cite{Yang62,Leggett04}, and the system is described by a macroscopic wavefunction of the Ginzburg-Landau equation \cite{Schrieffer71,Abrikosov75,Fetter03}.
After the broken U(1) symmetry, there exists collective modes, named Nambu-Anderson-Higgs-Goldstone  modes \cite{Higgs64,Goldstone62,Nambu59,Anderson58}, whose energy must vanish in the long wavelength limit to recover the symmetry of the system before the phase-transition.
In superconductors, these collective modes are difficult to observe, since electrons have charge, such that the phase fluctuation indues plasma oscillations \cite{Schrieffer71,Abrikosov75,Fetter03}, while similar collective modes were observed in cold atoms as Bogoliubov modes \cite{Ketterle02,Cornell02}.
In addition to these quasi-particle excitations, the beauty of BEC lies in the fact, that the entire system is described by a U(1) wavefunction to exhibit interferences as matter waves \cite{Ketterle02,Cornell02}.

We have recently revisited \cite{Saito20a,Saito20b,Saito20c,Saito20d,Saito20e,Saito21f,Saito22g,Saito22i,Saito23j} for the notion of the macroscopic coherence in the case of photons, emitted from a ubiquitously-available conventional laser source 
\cite{Stokes51,Poincare92,Jones41,Fano54,Baym69,Sakurai14,Max99,Jackson99,Yariv97,Gil16,Goldstein11,
Hecht17,Pedrotti07}.
Our hypothesis is that we might be able to describe the coherent state of photons as a macroscopic wavefunction to exhibit quantum mechanical behaviours at least for spin and orbital angular momentum states. 
We believe this hypothesis is highly non-trivial, since the state above lasing threshold is often considered to be classical.
But, do we have classical photons?
Photons are elementary particles, such that they are different from classical macroscopic objects. 
Then, in what sense can we consider the coherent state as classical electro-magnetic waves \cite{Max99,Jackson99}?
In particular, the calculus of Jones vectors for Stokes parameters $(S_1,S_2,S_3)$ on the Poincar\'e sphere is highly quantum mechanical, which is completely consistent with the 2-level systems for spin \cite{Stokes51,Poincare92,Jones41,Fano54,Baym69,Sakurai14,Max99,Jackson99,Yariv97,Gil16,Goldstein11,
Hecht17,Pedrotti07,Saito20a}. 
Why classical electro-magnetic waves behave quantum mechanically?
And perhaps more importantly, what makes the apparent quantum-mechanical behaviour to be classical, ceasing the coherence?

We have addressed these questions \cite{Saito20a,Saito20b,Saito20c,Saito20d,Saito20e,Saito21f,Saito22g,Saito22i,Saito23j}, and so far, it is consistent to consider coherent states for spin and orbital angular momentum are described by macroscopic wavefunctions.
Above lasing threshold for pumping \cite{Yariv97,Gil16,Goldstein11,Hecht17,Pedrotti07}, an SU(2) symmetry of photons (the special unitary group of 2-dimension), related to the rotational symmetry upon the propagation, is broken \cite{Saito20d} due to the natural selection of the lowest loss mode for stimulated emissions in a laser cavity.
For coherent photons from a laser source, there is no energy required to rotate the polarisation state to recover the SU(2) symmetry \cite{Saito20a,Saito20d}, since photons do not have charge.
There are no interaction between photons in a vacuum, and only weak non-linear interactions are expected in conventional materials \cite{Yariv97,Gil16,Goldstein11,Hecht17,Pedrotti07}.
Consequently, the SU(2) wavefunction describes the polarisation state, which is characterised by the spin expectation values shown on Poincar\'e sphere as Stokes parameters \cite{Stokes51,Poincare92,Jones41,Fano54,Baym69,Sakurai14,Max99,Jackson99,Yariv97,Gil16,Goldstein11,
Hecht17,Pedrotti07,Saito20a}. 

We have also examined orbital angular momentum \cite{Allen92,Padgett99,Milione11,Naidoo16,Liu17,Erhard18,Andrews21,Angelsky21,Agarwal99,Cisowski22},   
and found that the orbital angular momentum is a proper quantum-mechanical observable, which is well-defined in a waveguide or for a confined mode, predominantly propagating along a paraxial direction \cite{Saito20b,Saito20c,Saito23j}.
In particular, we have theoretically proposed to assigning 3 orthogonal states of left- and right vortexed states and no vortex state as SU(3) states \cite{Saito23j}, similar to colour charge for quarks in Quantum Chromo-Dynamics (QCD) \cite{Gell-Mann61,Gell-Mann64,Ne'eman61,Pfeifer03,Sakurai67,Georgi99,Weinberg05}.
We have employed a representation theory of the Lie algebra and the Lie group  \cite{Stubhaug02,Fulton04,Hall03,Pfeifer03,Dirac30}, and calculated expectation values for generators of rotations in the $\mathfrak{su}(3)$ Lie algebra, which form a hypersphere on SO(8) named as a Gell-Mann hypersphere \cite{Saito23j}.
The SU(3) state can also be characterised by 2 Poincar\'e spheres; one for orbital angular momentum and the other for hyperspin \cite{Saito23j} to represent couplings between a vortexed state and no-vortexed state.

Here, we experimentally show that such an SU(3) state is actually realised in a simple free-space optical experiment.
We have explored how topological charge is responsible to change from a vortexed state to no-vortexed state.
Moreover, we have also included spin degree of freedom together with orbital angular momentum to realise a singlet state and triplet states by using SU(4) states.

\section{Experimental set up}
In order to realise an arbitrary SU(3) state or even higher order photonic states, we need to prepare an SU(2) operator, since the rotations in SU(N) can be constructed by combining several SU(2) operators \cite{Stubhaug02,Fulton04,Hall03,Pfeifer03,Dirac30,Gell-Mann61,Gell-Mann64,Ne'eman61,Pfeifer03,Sakurai67,Georgi99,Weinberg05,Saito23j}.
For example, arbitrary SU(3) operations can be achieved by 2 sets of SU(2) operators, since the Lie algebra of $\mathfrak{su}(3)$ is rank 2 \cite{Pfeifer03,Georgi99,Saito23j}.
We have previously shown that an arbitrary SU(2) operation can be realised by a proposed Poincar\'e rotator \cite{Saito21f,Saito22g,Saito22h,Saito22i}, using fibre optic experiments.
Here, we have prepared a similar set-up for passive free space experiments.
This allows us to check the proof-of-concept easily, and it is also useful to extend the Poincar\'e rotator to allow coupling among orthogonal vortexed modes with spin-to-orbital conversion using vortex lenses \cite{Allen92,Padgett99,Milione11,Naidoo16,Liu17,Erhard18,Andrews21,Angelsky21,Agarwal99,Cisowski22,Golub07}.

\begin{figure}[h]
\begin{center}
\includegraphics[width=8cm]{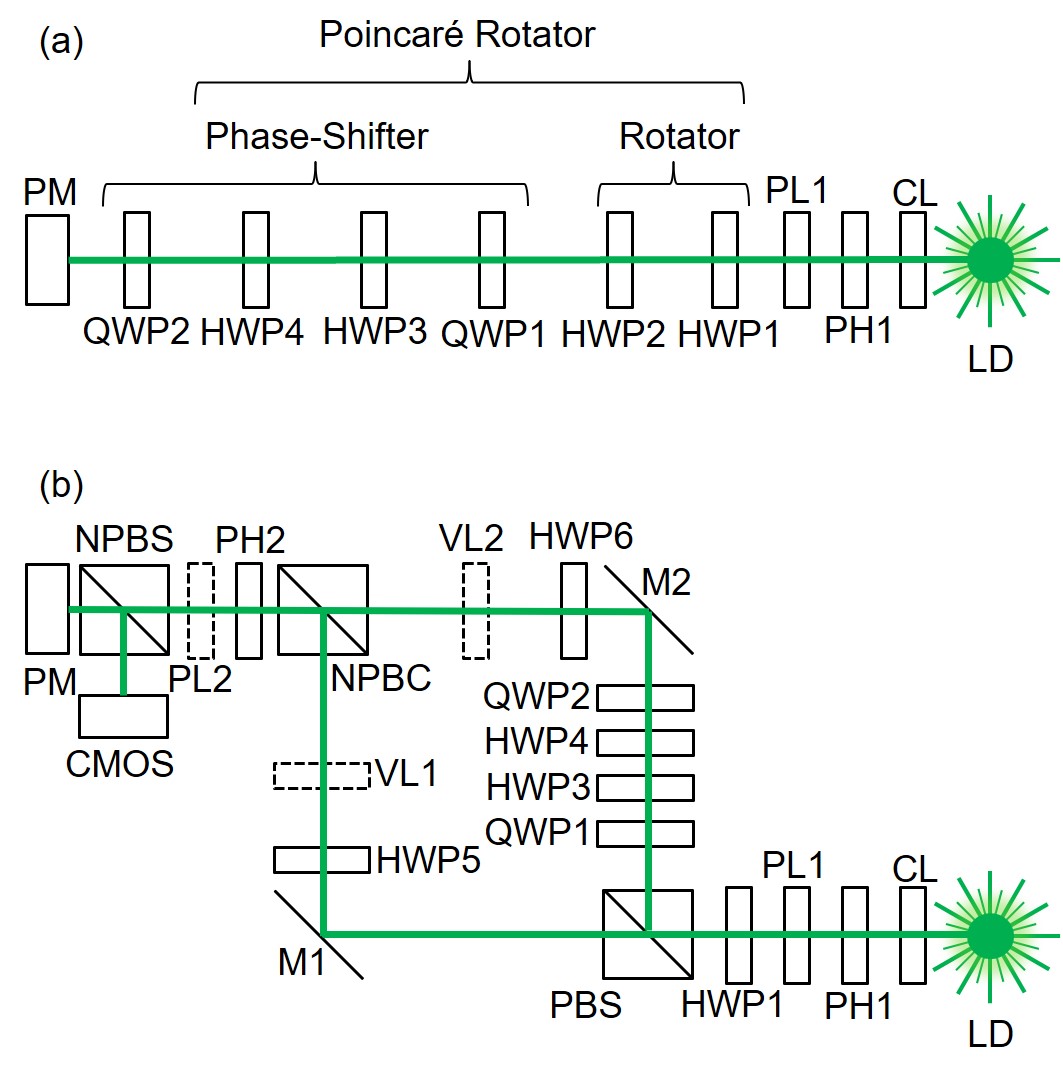}
\caption{
Experimental set up.
Abbreviations are as follows: 
LD: Laser Diode,  
CL: Collimator Lens,  
PH: Pin Hole,   
PL: Polariser,   
HWP: Half-Wave Plate,  
QWP: Quarter-Wave Plate, 
PBS: Polarisation Beam Splitter, 
NPBC: Non-Polarisation Beam Combiner,  
NPBS: Non-Polarisation Beam Splitter,   
M: Mirror,  
PM: Polarimeter,  
CMOS: Camera. 
(a) Poincar\'e rotator for polarisation. 
(b) Poincar\'e rotator for arbitrary orthogonal modes with and without a vortex.
VL1, VL2, and PL2 are shown by dotted lines, indicating that we have conducted experiments both with and without these optical components.
}
\end{center}
\end{figure}

\subsection{Poincar\'e rotator}
The proposed Poincare\'e rotator in free space \cite{Saito21f,Saito22g,Saito22h,Saito22i} is shown in Fig. 1 (a).
We have used a Diode-Pumped Solid State (DPSS) Laser-Diode (LD) with a wavelength of 532 nm, which was operated at the temperature of 20 $^{\circ}$C under the constant current of 120 mA with the output power of $\sim1$ mW.
The beam was collimated using a convex lens (a Collimator Lens, CL) with the focal length, $f$, of 100 mm.
The beam shape was refined by using a Pin Hole (PH) with a diameter of 200 ${\rm \mu}$m to make a Gaussian mode profile. 
The ray was passing through the polariser (PL) to make the polarisation state horizontal to an optical table under a vibration isolation.

Then, we have inserted into a Poincar\'e rotator \cite{Saito21f,Saito22g,Saito22h,Saito22i}, which comprises from a rotator and a phase-shifter.
The rotator is simply made of 2 successive operations of Half-Wave Plates (HWPs) \cite{Saito22g}.
The HWP2 was used upon rotating physically, such that its Fast Axis (FA) is aligned to a certain angle, measured from the horizontal direction.
It is well known that the rotated HWP works as a pseudo-rotator \cite{Gil16,Goldstein11}, which is actually a mirror reflection of the input polarisation state \cite{Saito20a,Saito22g}. 
The pseudo-rotator, associated with the corresponding mirror reflection, does not form a group \cite{Stubhaug02,Fulton04,Hall03,Pfeifer03,Georgi99}, since 2 successive operations of mirror simply correspond to an identity operation, while it is required to rotate twice the angle of the pseudo-rotation for establishing a group.
This issue is overcome, if we introduce additional HWP1 before operating by rotated HPW2 \cite{Saito20a,Saito22g}. 
The HWP1 was set its FA to align along the horizontal direction, which converts the pseudo-rotator to a pristine rotator to form a group for rotations \cite{Saito20a,Saito22g}.
Therefore, we can construct an rotator with an arbitrary rotation angle, determined by a physical rotation angle of HWP2 \cite{Saito20a,Saito22g}.
The rotator controls the relative amplitudes between horizontally-polarised and vertically-polarised states, such that it allows us to rotate the polarisation state on the Poincar\'e sphere along the azimuth direction in the $S_1$-$S_2$ plane.
The proper rotator corresponds to the U(1) $\cong$ SO(2) operation, which becomes a crucial component to realise rotations in higher SU(N).

It is important to be aware that the physical rotation angle of $\Delta \Psi$ for HPW2  corresponds to the 4 times of the rotation of the polarisation state on the Poincar\'e sphere, $4\Delta \Psi$, along the azimuthal direction \cite{Gil16,Goldstein11}.
The factor of 2 is coming from the mirror reflection of the complex electric fields, and another factor of 2 is coming from the difference of angles of SU(2) wavefunctions and its expectation value \cite{Saito20a,Saito22g}.
This factor of 4 is practically critical, since the uncertainty of physical rotation angle expands the deviation of the azimuthal angle on the Poincar\'e sphere from the expected value \cite{Gil16,Goldstein11,Saito20a,Saito22g}.

Then, the ray, rotated by a rotator, is inserted in a phase-shifter (Fig. 1 (a)), which rotates the polarisation state in the $S_2$-$S_3$ plane on the Poincar\'e sphere.
In order to realise a phase-shifter, we need to insert 2 Quarter-Wave Plates (QWPs) before and after the rotator.
The first QWP1 is aligned along the anti-diagonal direction, which is rotated $-45^{\circ}$ from the horizontal axis, seen from the detector side \cite{Saito20a}.
This makes the $S_2$-$S_3$ plane to the $S_1$-$S_2$ plane, such that we can rotate the polarisation state using a rotator, made of HWP3 and HWP4.
Finally, we can bring the rotated plane back to the original plane, by aligning the QWP2 to the diagonal direction, which is rotated $45^{\circ}$ from the horizontal axis \cite{Saito20a,Saito22g}, and the polarisation states were measured by a polarimeter (PM).

\subsection{Spin-to-orbit conversions and SU(2) operations}

We have realised an arbitrary rotation of spin for photons as polarisation states by the proposed Poincar\'e rotator \cite{Saito21f,Saito22g,Saito22h,Saito22i}.
We will use the Poincar\'e rotator as for a device to allow other SU(2) operations for photons with and without a vortex.
In order to realise a superposition state among orthogonal states with vortices, we have needed to extend the free space set-up to allow conversions of spin to orbital angular momentum using vortex lenses \cite{Allen92,Padgett99,Milione11,Naidoo16,Liu17,Erhard18,Andrews21,Angelsky21,Agarwal99,Cisowski22,Golub07}.

Figure 1 (b) shows the experimental set up, used in this work.
We have used the same DPSS LD at 532nm as the one used in Fig. 1 (a), and the ray is collimated by the same CL and  a PH1.
We used the PL1 to make the ray to the horizontally polarised state, and we used HWP1 for a rotator.
Here, we have removed HWP2, since there is no difference between a pseudo-rotator and a pristine rotator for a horizontally polarised state.
The advantage to make a pristine rotator configuration of Fig. 1 (b) is that the rotator works for any input, regardless of the polarisation states, including unknown polarisation states \cite{Saito20a}.
For our demonstration, it is not necessary to include HWP2, such that we have omitted to prepare.

Then, the ray was separated between horizontally polarised state and the vertically polarised state by a Polarisation Beam Splitter (PBS).
The PBS works as a converter from spin states to orbital paths, such that we can separate orthogonal polarisation states.
The ray with the horizontally polarised state is propagating to the left and reflected by the mirror (M1) to change the direction vertically.
We have inserted another HWP5, which can be rotated to switch from the horizontally polarised state to the vertically polarised state, by setting its FA to diagonally or anti-diagonally.
Or, we can also keep the horizontal polarisation state, simply by setting its FA horizontally or vertically.
This allows us to chose one of the polarisation state before combining again by a Non-Polarisation Beam Combiner (NPBC).
It was our option to include a vortex lens (VL1) or not, depending on whether we would like to convert the no-vortexed state to the vortexed state.
In our experiments, we tried both cases, such that VL1 in Fig. 1 (b) is shown by a dotted line.
When we included VL1, we intended to generate a left vortex, with orbital angular momentum of 1 per photon, in the unit of the Dirac constant of $\hbar$, whose chirality is defined by  seeing from the detector side \cite{Saito20a,Saito20b,Saito20c,Saito21f,Saito22g,Saito22h,Saito22i,Saito23j}.

For the path reflected vertically at the PBS, we have inserted a phase-shifter, using QWP1, HWP3, HWP4, and QWP2, exactly the same as the set-up, shown in Fig. 1 (a).
Here, only HWP4 was physically rotated to control the phase-shift, while QWP1, HWP3, and QWP2 were aligned to $-45^{\circ}$, 0, and $45^{\circ}$ directions, respectively, measured from the horizontal axis.
Note that the vertical polarisation state is not changed during the propagation through the phase-shifter operation by QWP1, HWP3, HWP4, and QWP2, since the phase-shifter rotates the polarisation state on the Poincar\'e sphere along the $S_1$ axis, and the vertically polarised state is located on the rotation axis of at $S_1=-1$.
Nevertheless, the phase-shift is actually accompanied to the vertically polarised state, which is observable in comparison with the other ray, going to M1.
Then, the ray after the phase-shifter is reflected at the M2, followed by the HWP6, which could be rotated to change the polarisation state.
We have an option to include VL2, as before, depending on whether we want to generate a vortex or not.
When we included VL2, we intended to generate a right vortex, with orbital angular momentum of -1.

The rays coming from M1 and M2 were combined by a NPBC, which was made of a polarisation independent splitter with the ratio of 50:50.
Therefore, the half of the intensity was lost, leaving to the vertical direction, which is not shown in Fig. 1 (b), for simplicity.
The combined ray was passing through PH2 with the same diameter of 200 ${\rm \mu}$m
 with PH1.
We have also examined the ray with and without PL2, which select a certain polarisation state, depending on the direction of the polariser.
Finally, the output ray was separated by a Non-Polarisation Beam Splitter (NPBS) to analyse both polarisation and far-field images, using a PM and a Complementary-Metal-Oxide-Semiconductor (CMOS) camera, respectively.

Here, we must be careful for the change of spin and orbital angular momentum states, upon applications of mirrors \cite{Max99,Jackson99,Yariv97,Gil16,Goldstein11,Hecht17,Pedrotti07,Saito20a}.
These operations effectively change the signs of axes of $x$ and $z$, while keeping the $y$ axis the same.
This could be easily understood, if we consider the total mirror reflection for the propagation against $z$, which will reflect the ray back to $-z$, such that the propagation direction is opposite and the right-handed $x$, $y$, and $z$, coordinate must be rotated $180^{\circ}$ along the $y$ axis.
In this coordinate, $S_2$ and $S_3$ should change sign upon the reflection, while $S_1$ should be the same \cite{Saito20a,Saito20e}.

Another choice of the frame after the reflection is to swap $x$ and $y$, while changing the sign of $z$, which is suitable to see the duality between complex conjugated states \cite{Saito20e}.
For the polarisation operation using $\mathfrak{su}(2)$ operators of Pauli matrices $\hat{\bm \sigma}=(\hat{\sigma}_1,\hat{\sigma}_2,\hat{\sigma}_3)$, this corresponds to use a complex conjugate representation by a mirror reflection of $\hat{\bm \sigma} \rightarrow \hat{\bar{\bm \sigma}}=-\hat{\bm \sigma}^{*}=(-\hat{\sigma}_1^{*},-\hat{\sigma}_2^{*},-\hat{\sigma}_3^{*})=(-\hat{\sigma}_1,\hat{\sigma}_2,-\hat{\sigma}_3)$ \cite{Georgi99,Saito20a,Saito20e}.
The same sign changes are also applicable to the vortexed states, since the mirrors change the coordinate and the chirality is changed.
Thus, for the SU(3) states, $\mathfrak{su}(3)$ operators of $\hat{\lambda}_i$ ($i=1$, $\cdots$, $8$) is reflected to be its complex conjugate $\hat{\bar{\lambda}}_i=-\hat{\lambda}_i^{*}$, and the wavefunction changes sign, accordingly \cite{Saito20a,Saito23j}. 

In our set-up, shown on Fig. 1 (b), the ray, going to the M1 with the horizontally polarised state, is reflected twice at M1 and NPBC to enter the PM, such that the twice of mirror operations will not change the polarisation state.
Similarly, for the ray, going to M2 with the vertically polarised state, is reflected at the PBS and M2, such that no change of the polarisation state is expected for the ray entering into the PM.
Here, it is very important to make sure that the number of mirror reflections should be the same for both paths.
It is also allowed to have an even number of difference among different paths, but an odd number will mix rays, whose polarisation states are defined in a different coordinate and the analysis of polarisation states will be more difficult.
For the ray going to a CMOS camera, we have an additional 1 more reflection at the NPBS, such that the polarisation state at the CMOS camera is the mirror reflected state of that at the PM.
For the left vortex generated at VL1 will be kept left at the CMOS camera, since 2 mirror reflections are expected at the NPBC and the NPBS.
On the other hand, only one reflection at NPBS is required for the vortexed state generated at VL2 to entering into the CMOS camera.
Therefore, we just need to generate the same left vortex at the VL2 for expecting the vortex to be the right vortex at the CMOS camera.
This is absolutely fine, since the left vortex generated at VL1 will become the right vortex after the reflection at NPBC, such that after passing through NPBC, both left and right vortexed states are combined to form 1 ray.

\section{Experimental Results and Discussions}

\subsection{Polarisation controlled by Poincar\'e rotator}

\begin{figure}[h]
\begin{center}
\includegraphics[width=8cm]{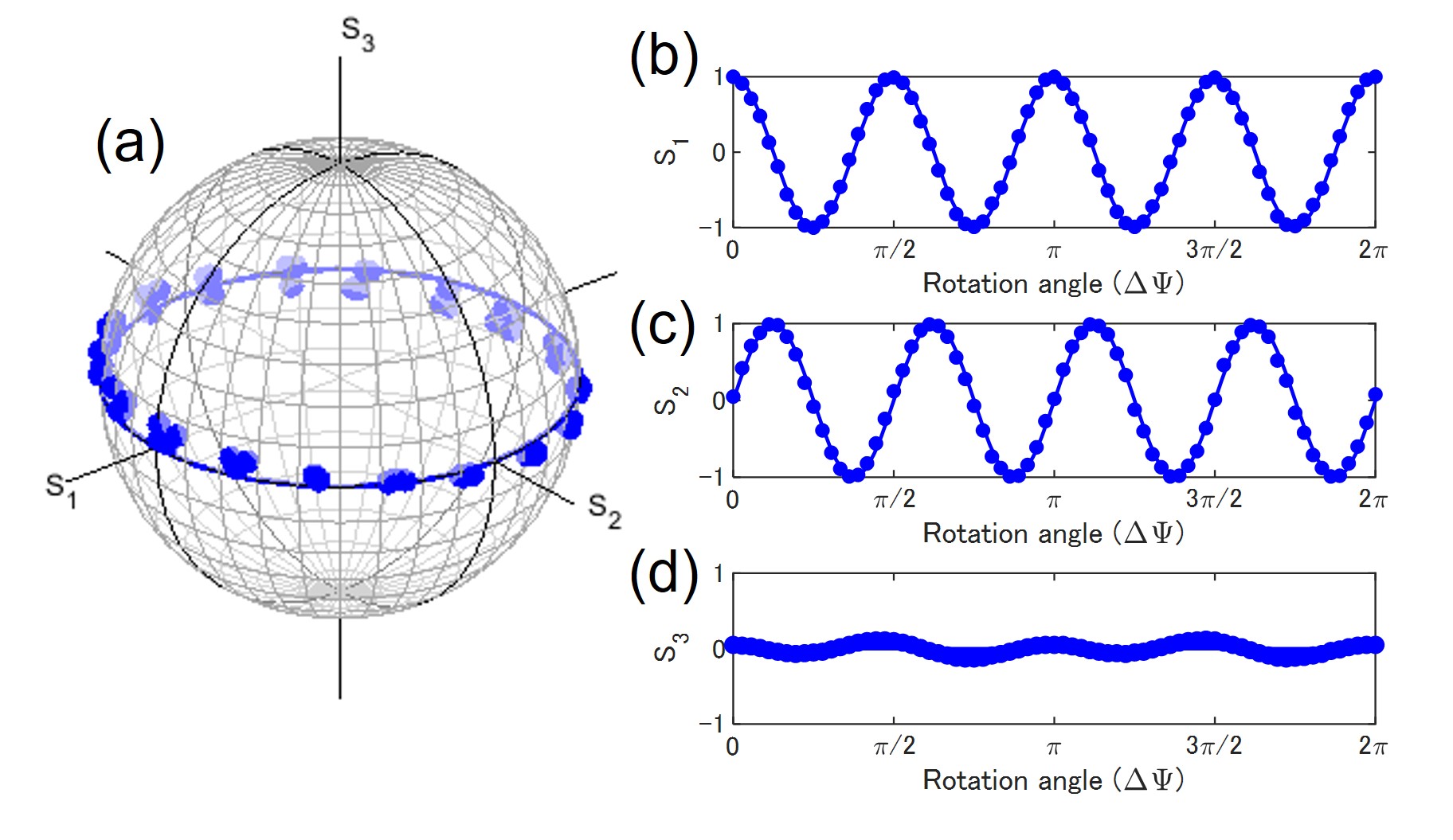}
\caption{
Rotator operation of the Poincar\'e rotator, shown on (a) Poincar\'e sphere. 
Stokes parameters of (b) $S_1$, (c) $S_2$, and (d) $S_3$ are also shown.
We have inserted a horizontally polarised ray into the Poincar\'e rotator, and we have fixed the phase-shift to allow rotation of the polarisation states along the equator of the Poincar\'e sphere.
At the zero physical rotation angle, $\Delta \Psi$, the polarisation state was not changed, while the polarisation state was rotated along the anti-clock-wise direction (left direction), seen from the top of the $S_3$ axis.
The polarisation states were rotated 4 times upon the physical rotation of the half-wave-plates, as expected from the SU(2) theory for spin states of coherent photons.
}
\end{center}
\end{figure}

First, we have used the experimental set-up of Fig. 1 (a), and confirmed the expected operations as a passive Poincar\'e rotator using a green LD.
We show the rotator operation of the Poincar\'e operator along the equator of the Poincar\'e sphere.
In this experiment, we have set the physical rotation angle ($\Delta \Psi$) of HWP2 at every 5$^{\circ}$, while other plates were kept fixed.
We confirmed that the polarisation states were rotated upon rotating HWP2, as expected from the SU(2) theory for spin states of photons \cite{Saito20a,Saito23j}.

In fact, we have found the fundamental relationship between the rotation ($\theta_a$) of the SU(N) state, and and the rotation of the expectation value for the generator of the rotation in SO($N^2-1$) \cite{Saito23j}. 
This means that the SU(N) wavefunction is rotated as
\begin{eqnarray}
|{\rm F} \rangle
&=&
{\rm e}^{-i \hat{X}_a \theta_a}
|{\rm I} \rangle,
\end{eqnarray}
where $|{\rm I} \rangle$ and $|{\rm F} \rangle$ refers to the initial and the final states, and $\hat{X}_a$ stands for the generator of the rotation along the $a$ axis.
Then, the expectation value is obtained as  
\begin{eqnarray}
\langle 
\hat{X}_b
\rangle_{\rm F}
&=&
\sum_c
\left( 
{\rm e}^{- \hat{F}_{a} \theta_a}
\right)_{bc}
\langle \hat{X}_c \rangle_{\rm I}, 
\end{eqnarray}
where $\langle \hat{X}_b \rangle_{\rm F}$ and $\langle \hat{X}_c \rangle_{\rm I}$ are expectation values in the initial and final states along $b$ and $c$ axes, respectively, and $\hat{F}_{a}$ is an adjoint operator, whose matrix element, $(\hat{F}_a)_{bc}=f_{abc}$ is the structure constant of the commutation relationship, $ [ \hat{X}_a,\hat{X}_b ] =i\sum_{c} f_{abc} \hat{X}_c$.

In the present example between SU(2) and SO(3), using the orthogonal bases of horizontal and vertical states \cite{Saito20a}, the generators become spin operators as $(\hat{X}_1,\hat{X}_2,\hat{X}_3)=(\hat{s}_1,\hat{s}_2,\hat{s}_3)/\hbar=(\hat{\sigma}_3,\hat{\sigma}_1,\hat{\sigma}_2)/2$, and the structure constant is given by the totally anti-symmetric tensor $f_{abc}=\epsilon_{abc}$. 

Consequently, our formula for the rotator, provided by setting $a=3$ for the rotation along the $S_3$ axis, gives
\begin{eqnarray}
\left (
  \begin{array}{c}
S_1
\\
S_2
\\
S_3
  \end{array}
\right)
&=&
\left (
  \begin{array}{ccc}
\cos \theta_3   &   -\sin \theta_3 & 0
\\
\sin \theta_3   & \cos \theta_3  &0
\\
0 & 0 & 1
  \end{array}
\right)
\left (
  \begin{array}{c}
1
\\
0
\\
0
  \end{array}
\right)
\\
&=&
\left (
  \begin{array}{c}
\cos \theta_3 
\\
\sin \theta_3 
\\
0
  \end{array}
\right),
\end{eqnarray}
where the rotation angle on the Poincar\'e sphere is given by $\theta_3=4 \Delta \Psi$, since the rotated HWP changes the angle of the complex electric fields twice the angle of the physical rotation due to the nature of the mirror reflection \cite{Saito20a}, and another factor of 2 is coming from the difference between complex electric fields in the real space and its expectation value \cite{Saito20a}.
The latter difference is understood by noting horizontal and vertical states are orthogonal in the real space with the angle of $90^{\circ}$, while they are located in the opposite side of the same $S_1$ axis with the angle of $180^{\circ}$ on the Poincar\'e sphere. 
Actually, we confirmed the 4-times rotations over the equator in our experiments (Fig. 2)

\begin{figure}[h]
\begin{center}
\includegraphics[width=8cm]{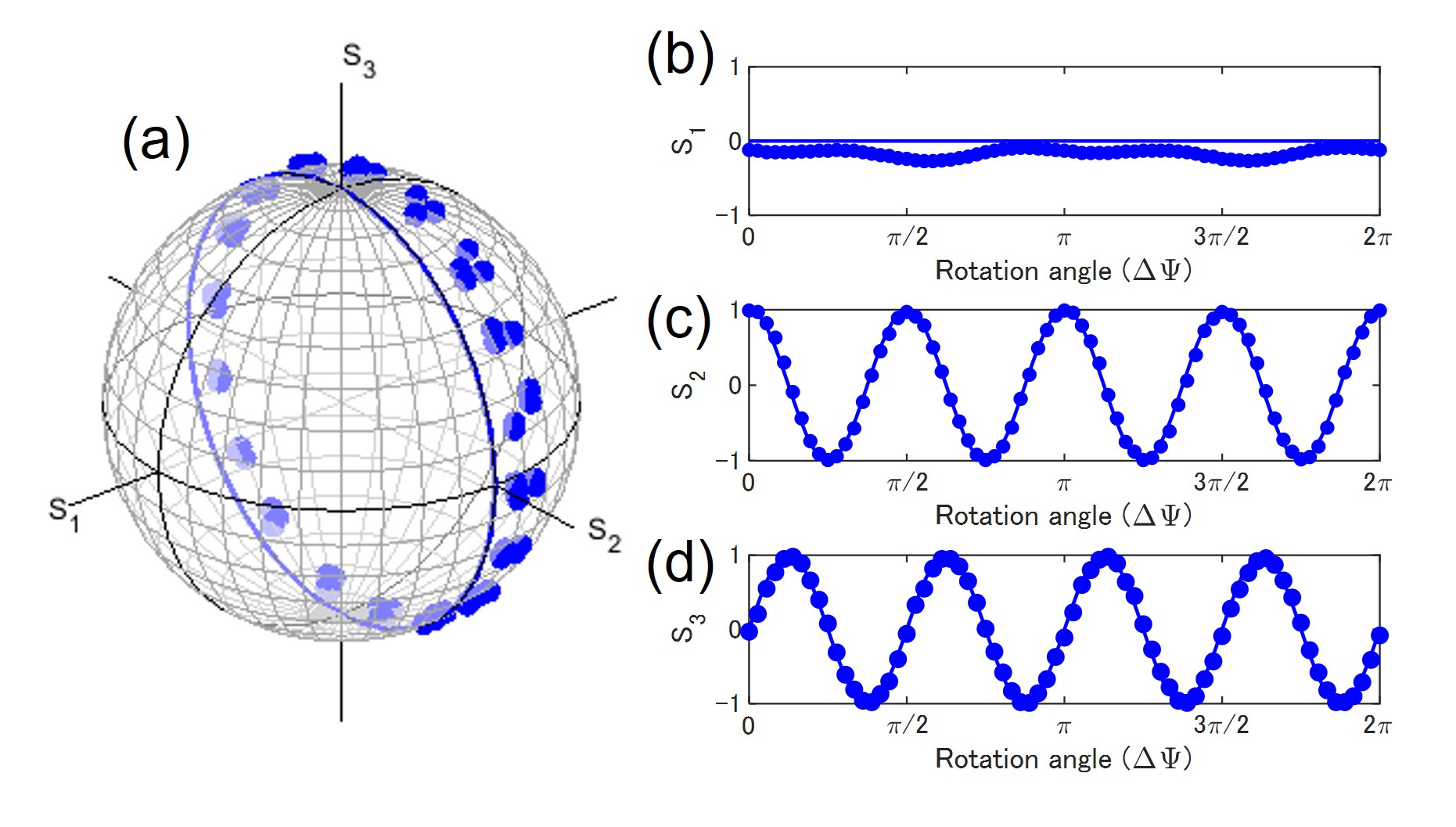}
\caption{
Phase-Shifter operation of the Poincar\'e rotator, as shown on (a) Poincar\'e sphere and by Stokes parameters of (b) $S_1$, (c) $S_2$, and (d) $S_3$.
We found deviations from theoretical lines, which are considered to be within the expected errors of $\sim 10^{\circ}$.
}
\end{center}
\end{figure}

Similarly, we have also confirmed the phase-shifter operation, as shown in Fig. 3.
Here, we have rotated HWP2 to align its FA to $\pi/8=22.5^{\circ}$ form the horizontal direction to convert the horizontal input to the diagonally polarised state, located at $S_2=1$ before entering into the phase-shifter,
Then, we have physically rotated HWP4, while other waveplates were kept fixed.
In this case, the phase-shifter rotated the polarisation state on the Poincar\'e sphere in the $S_2$-$S_3$ plane, as we expected from the SU(2)-SO(3) relationship by setting $a=1$ for the phase-shift, corresponding to the rotation along the $S_1$ axis, as 
\begin{eqnarray}
\left (
  \begin{array}{c}
S_1
\\
S_2
\\
S_3
  \end{array}
\right)
&=&
\left (
  \begin{array}{ccc}
1   &   0 & 0
\\
0  & \cos \theta_1  & -\sin \theta_1
\\
0 & \sin \theta_1  & \cos \theta_1
  \end{array}
\right)
\left (
  \begin{array}{c}
0
\\
1
\\
0
  \end{array}
\right)
\\
&=&
\left (
  \begin{array}{c}
0
\\
\cos \theta_1 
\\
\sin \theta_1 
  \end{array}
\right).
\end{eqnarray}

The experimental data is consistent with this theoretical expectations, shown by lines.
However, we found remarkable deviations, corresponding to the angles of $\sim10^{\circ}$, which are expected from the waveplates used in this experiments.
Our waveplates have deviations of around a few degree at the wavelength of 532nm, and the physical rotation of wave-plates would also induce uncertainty of $\sim 1^{^\circ}$
Moreover, the rotation angle on the Poincar\'e sphere is 4 times larger than the physical rotation angle, such that the errors of the angles also increase 4 times \cite{Gil16,Goldstein11,Saito20a}.

\begin{figure}[h]
\begin{center}
\includegraphics[width=8cm]{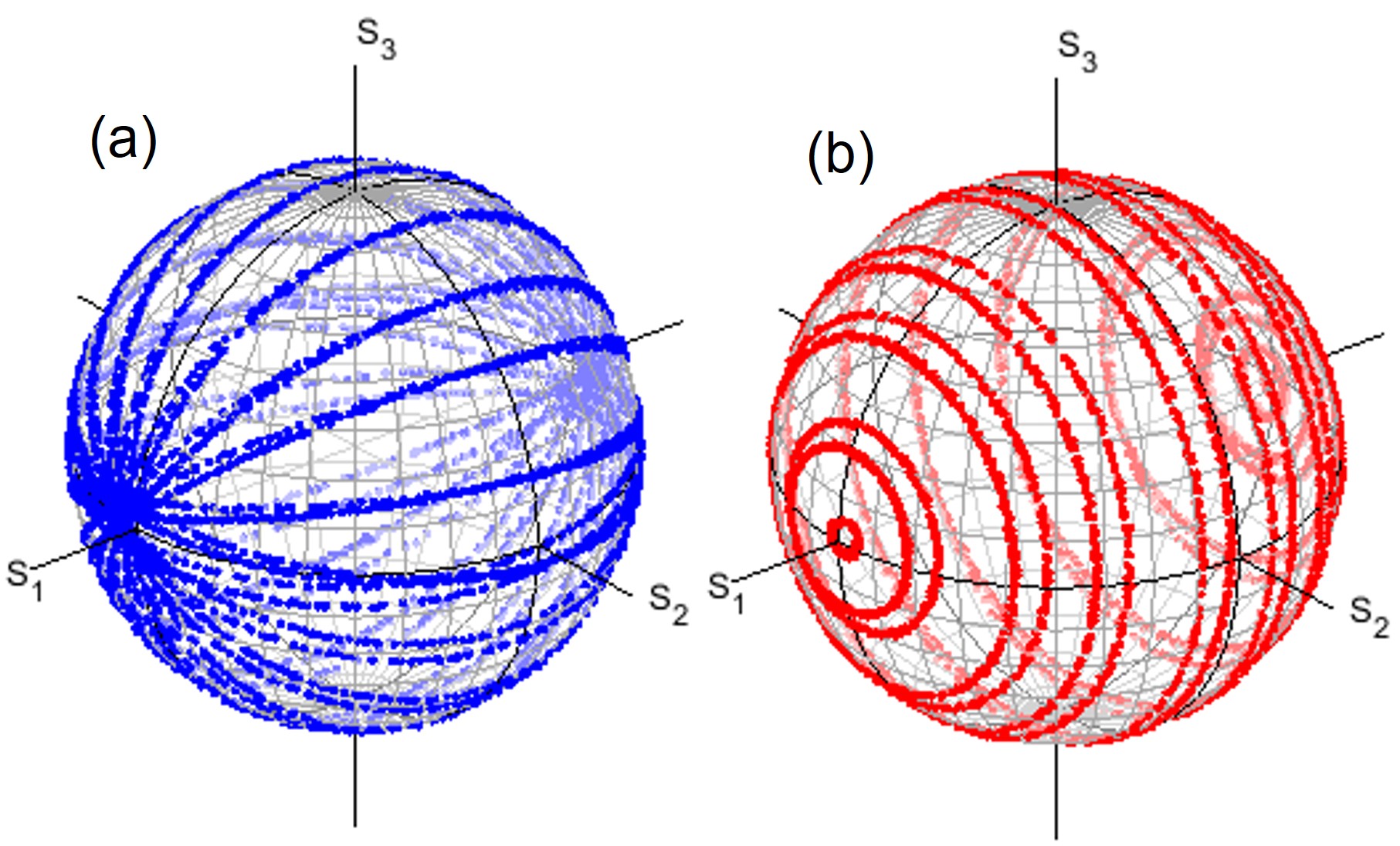}
\caption{
Poincar\'e rotator operations as (a) a rotator and (b) a phase-shifter.
The Stokes parameters were obtained by rotating corresponding wave-plates for 30s, which were recorded by the polarimeter.
}
\end{center}
\end{figure}

We have also operated the Poincar\'e rotator by using both rotator and phase-shifter angles, as shown in Fig. 4.
For the rotator operation of Fig. 4 (a), we have fixed the physical rotation angle for the phase-shifter at every $5^{\circ}$, while we have continuously rotated HWP2 for 30s by hands and recorded Stokes parameters.
We confirmed expected rotations for horizontal-vertical bases \cite{Saito20a}, since the north and the south poles are located at $S_1=1$ and $S_1=-1$, respectively, for fixed phase-shifts, whose trajectories are similar to the longitude for the earth.
On the other hand, we have also confirmed the phase-shifter rotations upon the fixed rotator angles.
We have fixed the physical rotation angle of HWP2 at every $5^{\circ}$, while we have continuously rotated HWP4 by hands for 30s and recorded by the PM.
As shown in Fig. 4 (b), the rotation plane is always parallel to the $S_2$-$S_3$ plane, since the phases between horizontal and vertical states are defined without changing the amplitudes, which corresponds to keep the $S_1$ value, when we used the horizontal and vertical states as orthogonal bases \cite{Saito20a}.
The trajectory in this case is similar to the latitude for the earth.

\subsection{Quantum coherency as SU(2) states}

Next, we have proceeded to examine the experimental set-up, shown in Fig. 1 (b), without VL1, VL2, and PL2 to confirm the operation as Poincar\'e rotator for polarisation, similar to the one, shown in Fig. 1 (a).
A practical challenge was the alignment.
If 2 rays were separated, going to different directions, they are completely different orthogonal modes, since the momentum for each ray is different.
Even if it looked aligned, monitored by a CMOS camera, modes of the rays must be perfectly matched to expect the quantum coherency as a single mode with the SU(2) symmetry as polarisation.
Without the coherency, a ray would not be rotated its polarisation state by the generator of rotations for the phase and the amplitude.
The measure of the quantum coherency (Fig. 5) is given by the degree of polarisation 
\begin{eqnarray}
Q
&=&
\frac{\sqrt{S_x^2+S_y^2+S_z^2}}{S_0}
,
\end{eqnarray}
where ${\bf S}=(S_x,S_y,S_z)$ are spin expectation values for coherent photons and $S_0$ is the magnitude of the total spin density for photons \cite{Stokes51,Poincare92,Jones41,Fano54,Baym69,Sakurai14,Max99,Jackson99,Yariv97,Gil16,Goldstein11,
Hecht17,Pedrotti07,Bjork10,Saito20a}.
We have previously shown that the Stokes parameters are actually given by normalising the spin expectation values
\begin{eqnarray}
\left (
  \begin{array}{c}
S_x
\\
S_y
\\
S_z
  \end{array}
\right)
&=&
\hbar N
\left (
  \begin{array}{c}
\sin \theta \cos \phi
\\
\sin \theta \sin \phi
\\
\cos \theta 
  \end{array}
\right)
\end{eqnarray}
to have the radius of the unit length, where $\theta$ is the polar angle, measured from the $S_3$ axis, $\phi$ is the azimuth in the $S_1$-$S_2$ plane, measured from the $S_1$ axis, and $N$ is the number of coherent photons per second (or per bit), passing through the cross section, perpendicular to the direction of the propagation.
If we would like to convert it in horizontal-vertical bases with poles located in the $S_1$ axis, it becomes
\begin{eqnarray}
\left (
  \begin{array}{c}
S_x
\\
S_y
\\
S_z
  \end{array}
\right)
&=&
\hbar N
\left (
  \begin{array}{c}
\cos (2 \alpha)
\\
\sin (2 \alpha) \cos \delta
\\
\sin (2 \alpha) \sin \delta
  \end{array}
\right),
\end{eqnarray}
where $alpha$ is the auxiliary angle to define the complex electric fields as $({\mathcal E}_x,{\mathcal E}_y)=E_0({\rm e}^{-i\delta/2}\cos\alpha,{\rm e}^{i\delta/2}\sin\alpha)$ with the $U(1)$ wavefunction for the orbital is $E_0$, and $\delta$ is the phase, defined between horizontal and vertical components.
On the other hand, the magnitude of the spin is given by
\begin{eqnarray}
S_0
=
\hbar N_0,
\end{eqnarray}
where $N_0$ is the number of total photons per second (or per bit), passing through the cross section, perpendicular to the direction of the propagation. 
Therefore, the degree of polarisation simply means the ratio between the total and the coherent numbers of photons, and we obtain 
\begin{eqnarray}
Q
=
\frac{N}{N_0}.
\end{eqnarray}

\begin{figure}[h]
\begin{center}
\includegraphics[width=8cm]{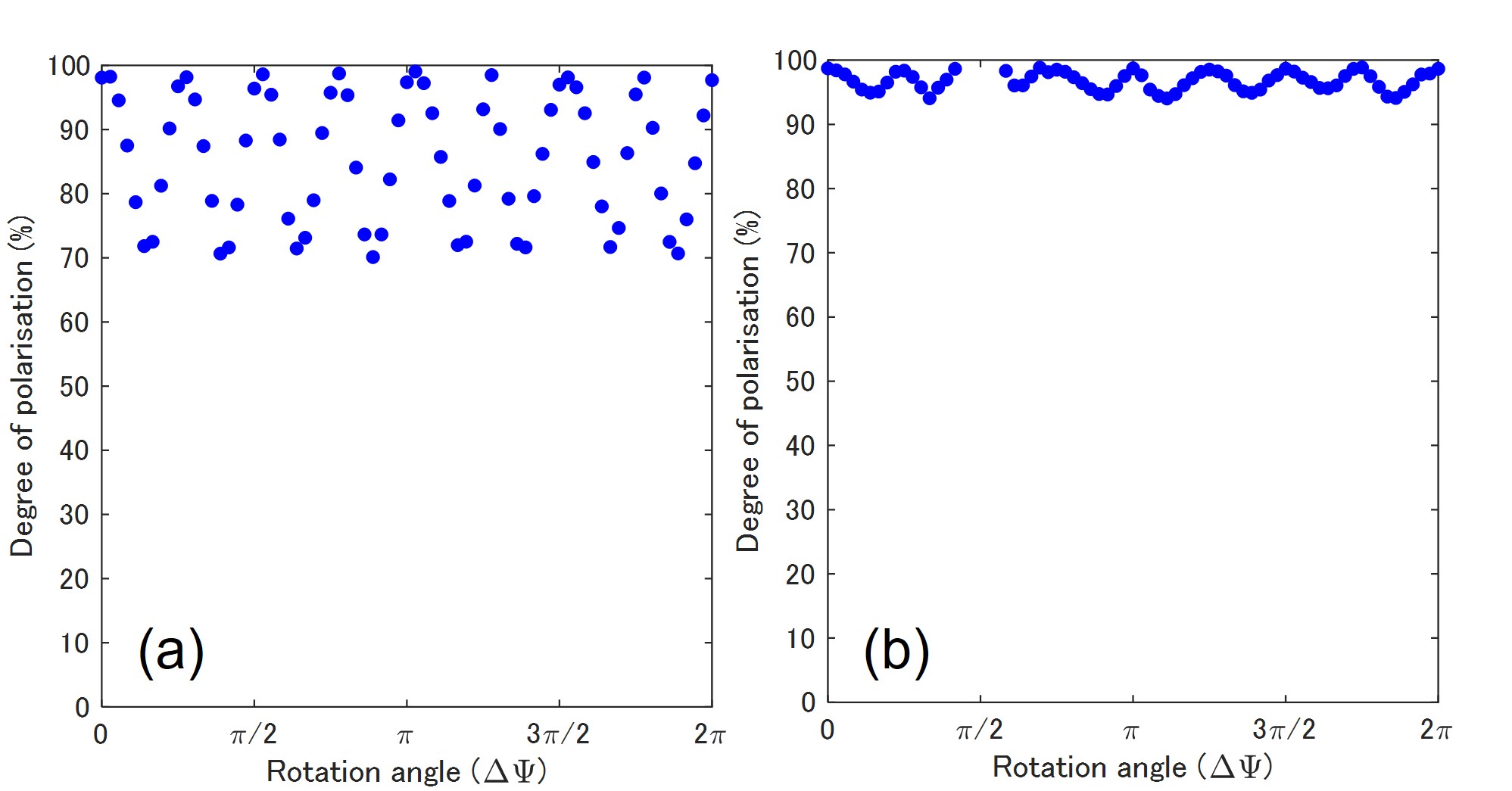}
\caption{
Degree of polarisation for Poincar\'e rotator (a) before and (b) after intensive alignment efforts.
}
\end{center}
\end{figure}

In our experiments for the set-up, shown in Fig. 1 (a), we  confirmed $Q$ above 99 \%.
On the other hand, the typical values of $Q$ before intensive efforts on alignments were below 1\%, especially worse, if we did not introduce PH2 to remove contributions of miss-aligned components.
Our beam divergence angle was around $\sim 1$ mrad, which expands our beam profile to be the diameter of $\sim 1$ mm at the entrance of PH2.
By taking only the central contribution, we could improve $Q$, as shown in Fig. 5, at the expense of reduced output power down to the order of $\sim 10$ ${\rm \mu}$W. 
There was no spacial technique, introduced for alignments, and only patience helped to improve $Q$, as is always truce for any optical experiments. 
We have intended to secure $Q>90$ \% in this work (Fig. 5 (b)), since we also expect similar amount of deviations upon rotating wave-plates (Fig. 4), such that our results below should be understood only qualitatively to see the trends, rather than full quantitative controls over the higher order Poincar\'e sphere \cite{Allen92,Padgett99,Milione11,Naidoo16,Liu17,Erhard18,Andrews21,Angelsky21,Agarwal99,Cisowski22,Golub07,Saito23j}.
After the optimisation, we have confirmed the successful operations of Poincar\'e rotator in the free space to obtain similar changes of polarisation states to those shown in Fig. 4 by rotating HWP1 and HWP4, shown in Fig. 1 (b).


Away from practical efforts and experimental designs, we think there is a hint, possibly, to understand the emergence of classical behaviours to loose the quantum coherence.
Before splitting orthogonal polarisation modes at the PBS, shown in Fig. 1 (b), two orthogonal modes were perfectly aligned and we could control amplitudes and the phase by Poincar\'e rotator, shown in Fig. 1(a) to switch from the horizontal state to the vertical state, or {\it vice versa} and to change the phase without affecting the coherency.
Once we split them at PBS, it is impossible to exchange the amplitudes any more as far as two rays were separated.
In that sense, the rays lost the macroscopic quantum coherency of SU(2) states, and 2 rays could be treated as classical independent rays without the SU(2) symmetry between the rays.
Nevertheless, the original amplitudes and the phase are memorised in each rays, as far as they have not affected by another loss mechanisms.
Therefore, we could control the phase by the phase-shifter, and we could also recover the quantum coherency as SU(2) states, after combining at the NPBC (Fig. 5).

We think this decoherence and its recovery process are different from the scenario, that the dissipative coupling to the environment induces the decoherence to the classical system to cease the quantum coherence \cite{Caldeira81,Leggett85,Leggett95}.
According to an instanton method to use an imaginary time path-integral \cite{Coleman77,Coleman88}, one can figure out a classical path even for the Macroscopic-Quantum Tunnelling (MQT) process of a false vacuum and ultimately for the Macroscopic-Quantum Coherence (MQC).
In reality, MQT and MQC are difficult to observe and it was required to understand why macroscopic valuables are difficult to exhibit quantum behaviours \cite{Caldeira81,Leggett85,Leggett95,Frowis18}.
In the case of a macroscopic quantum valuable, however, the dissipative coupling to the environment provides a frictional term, related to the history of the path, for reducing the quantum motion.
Following this scenario, it is encouraged to explore smaller systems for observing Schr\"odinger's cat-like phenomena
\cite{Caldeira81,Leggett85,Leggett95,Frowis18} to understand the emergence of the classical mechanics out of quantum mechanics more deeply.
In our present experiments, we just employed a conventional laser, such that photons are macroscopically degenerate to occupy the single mode like BEC, and there is no conceptual difficulty in the macroscopic coherence like Schr\"odinger's cat.
The photons are bosons and they could occupy the same state, and the whole system is simply described by the sole wavefunction, simply because photons are in the same state, which is described by the SU(2) wavefunction for the polarisation.
Nevertheless, photons could be the superposition state among 2 orthogonal polarisation states, which can loose the decoherence upon splitting, and, more importantly, the coherence can be recovered by the precise alignment at the expense of the loss in power.
It is less exciting compared with Schr\"odinger's cat, but still coherent photons are considered to be in a macroscopically-degenerate quantum state, and we can explore its SU(2) state, using a standard quantum-mechanical procedure \cite{Dirac30,Baym69,Sakurai14,Sakurai67,Weinberg05} and a representation theory of Lie algebra and Lie group \cite{Stubhaug02,Fulton04,Hall03,Pfeifer03,Georgi99,Saito23j}.

\subsection{Poincar\'e rotator for orbital angular momentum}

\begin{figure}[h]
\begin{center}
\includegraphics[width=8cm]{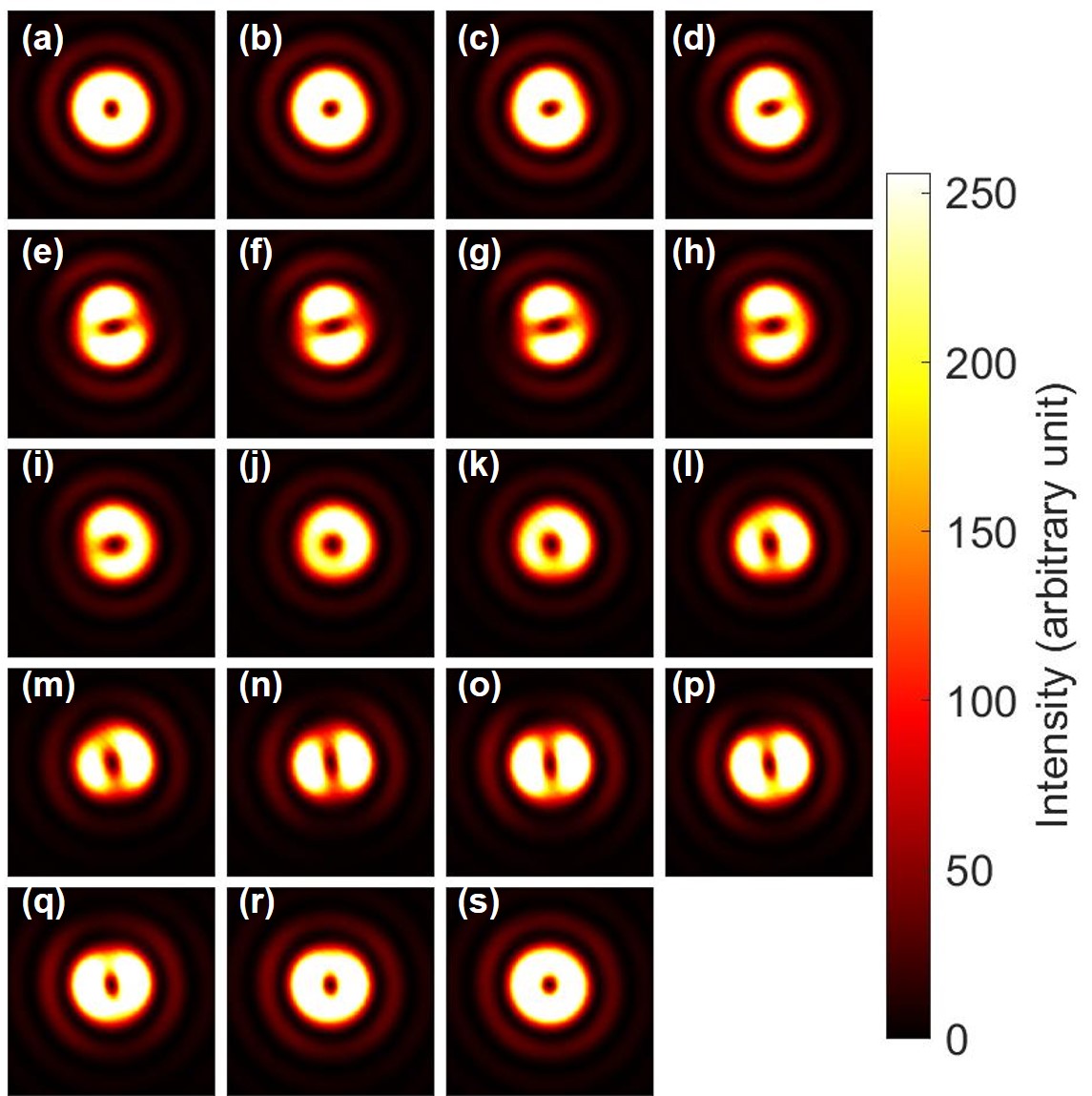}
\caption{
Rotator operation of Poincar\'e rotator for the orbital angular momentum.
The half wave plate was physically rotated from (a) 0$^{\circ}$ to (s) 90$^{\circ}$ with a step of 5$^{\circ}$, which is enough to rotate from (a) the left vortex, via (j) the right vortex, back to (s) the left vortex, for circulating the Poincar\'e sphere.
Dipole images were seen for (e)-(h) and for (m)-(p), which are orthogonal each other for the direction of the dipole.
}
\end{center}
\end{figure}

Next, we proceed to apply our simple experimental set-up for realising SU(2) superposition state for orbital angular momentum \cite{Allen92,Padgett99,Milione11,Naidoo16,Liu17,Erhard18,Andrews21,Angelsky21,Agarwal99,Cisowski22,Saito23j}.
Here, we have introduced both VL1 and VL2 to induce both left- and right-vortexed states to realise an arbitrary superposition state by our Poincar\'e rotator.
We have not inserted PL2, for the moment, and HWP5 was set to align its FA to the $45^{\circ}$-rotated direction, measured from the horizontal axis.
This corresponds to convert the horizontally polarised state for the ray, passing to M1, to the vertically polarised state.
On the other hand, the FA of HWP6 was aligned to be horizontal, keeping the vertically polarised state for the ray, passing to M2.
Therefore, both rays are in the vertically polarised state, such that we have converted the spin degree of freedom to the orbital angular momentum for controlling amplitudes and the phase of left- and right-vortexed states.
We have generated the same magnitude of the topological charge of 1 for both left- and right-vortexed states.
The amplitudes between left- and right-vortexed states were controlled by HWP1 to change the splitting ratio at PBS.
On the other hand, the phase was controlled by the phase-shifter, made of QWP1, HWP3, HWP4, and QWP2.
It must be mentioned that we have not controlled a global phase, solely by the difference of physical length upon propagation.
Nowadays, it is not impossible to control the distance by using a stepping motor to control the distance with a resolution of the order of a few nm per step.
In stead, we controlled the phase by the phase-shifter, such that the origin of the phase is not coincide with the physical rotation angle of HWP4.

As shown in Fig. 6, we observed famous expectations of SU(2) states with orbital angular momentum \cite{Allen92,Padgett99,Milione11,Naidoo16,Liu17,Erhard18,Andrews21,Angelsky21,Agarwal99,Cisowski22,Saito23j}, simply by   
rotating HWP1 as a rotator.
First, we confirmed a doughnut-like image of vortexed states at Fig. 6 (a), (j), and (s), as expected for topological charge, stored at the centre of the mode \cite{Allen92,Padgett99,Milione11,Naidoo16,Liu17,Erhard18,Andrews21,Angelsky21,Agarwal99,Cisowski22,Saito20b,Saito20c}.
We could also clearly confirm the dipole structures, realised by the superposition of left- and right-vortexed states, as shown in Fig. 6 (e)-(h) and (m)-(p), which are orthogonal to each other.
This means that superposition states of orthogonal orbital angular momentum states are actually formed, and SU(2) states of the orbital angular momentum can be shown on a Poincar\'e sphere, and our rotator successfully controlled the change of states over the equator of the sphere.
We needed to rotate HWP1 just $90^{\circ}$ to realise 1 rotation in the SO(3) space, which was exactly the same as the rotation angle, required for polarisation states.

\begin{figure}[h]
\begin{center}
\includegraphics[width=8cm]{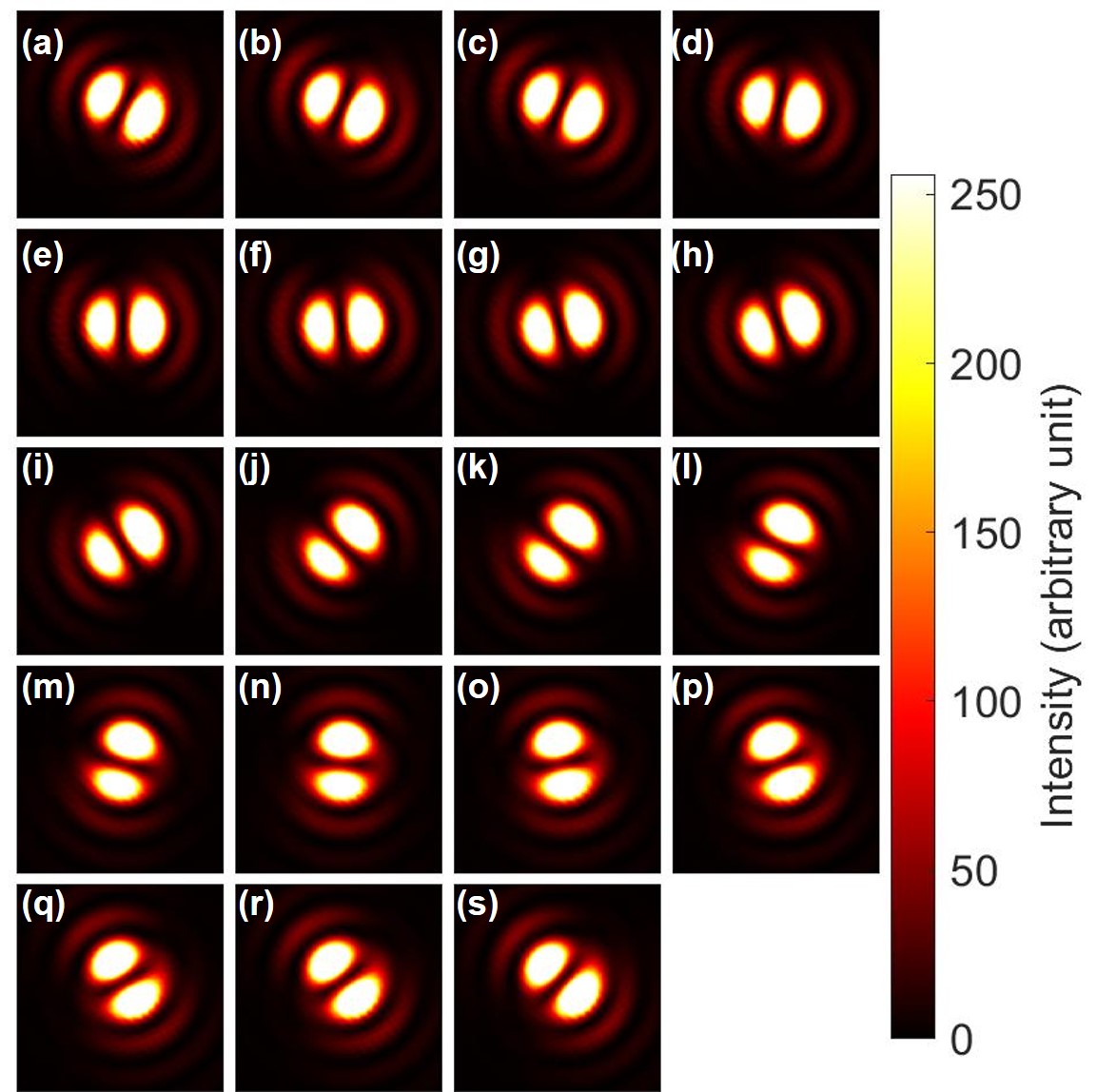}
\caption{
Phase-shifter operation of Poinca\'e rotator for orbital angular momentum.
The half-wave plate was physically rotated from (a) 0$^{\circ}$ to (s) 90$^{\circ}$ with a step of 5$^{\circ}$, to realise the phase-shift up to $\pi$.
Note that the dipoles were rotated upon the rotations the half-wave plate without changing the shape.
The physical rotation of 45$^{\circ}$ is required to rotate the dipole for 90$^{\circ}$, which corresponds to change the state orthogonal to each other, corresponding to the 180$^{\circ}$-rotation on the Poincar\'e sphere.
}
\end{center}
\end{figure}

Next, we have confirmed the phase-shifter operation of the Poincar\'e rotator for orbital angular momentum states.
We have set HWP1 to align its FA to $22.5^{\circ}$ from the horizontal direction, and we observed the dipole by the CMOS camera.
This allows to split the ray into 2 rays with the 50:50 splitting ratio at the PBS.
Then, we have rotated HWP4 to control the phase between left- and right-vortexed states.
As shown in Fig. 7, we observed the rotation of the dipole, upon rotating HWP4.
In other words, we could identify the phase, from images of the dipole.
Here, we must be careful about the phase.

As explained above, we have not controlled the global phase, such that the orientation of the dipole at the zero-physical rotation of HWP4, as defined for the alignment of its FA to the horizontal direction, (Fig. 7 (a)) was not properly designed to align to a specific direction.
We can set the phase, after observing the dipole image, to compensate the global phase, if we want.

The other interesting aspect on the phase is the relationship between the amount of rotation and the dipole image.
If we carefully track the images from Fig. 7 (a) to Fig. 7 (s), one can identify the dipole is just rotating $180^{\circ}$ upon the physical rotation of $90^{\circ}$ for HWP4. 
This will bring one of two bright regions in the dipole image of Fig. 7 (a), located at the right-down region, to the other place of the left-up region in the dipole image of Fig. 7 (s).
In the images, we could only recognise the intensities, but it is widely known that the phase of the dipole is opposite among 2  bright high-intensity regions.
This means that the phase is opposite between the dipole of Fig. 7 (a) and the dipole of Fig. 7 (s), while average angular momentum defined on the SO(3) Poincar\'e sphere must be the same.
This is indeed the manifestation of the two-fold coverage of SU(2) for SO(3), as mathematically known as ${\rm SU(2)}/{\mathbb S}^0 \cong {\rm SU(2)}/{\mathbb Z}_2 \cong {\rm SO(3)}$, where ${\mathbb S}^0= \{ -1, 1 \}$ and ${\mathbb Z}_2= \{ 0, 1 \}$ \cite{Fulton04,Hall03,Pfeifer03,Georgi99,Saito23j}.
Physically, expectation values often dismiss the geometrical phase information, such that we must be careful for their important roles \cite{Pancharatnam56,Berry84,Hamazaki06,Bliokh09,Cisowski22}.

\subsection{Interference fringes}

\begin{figure}[h]
\begin{center}
\includegraphics[width=8cm]{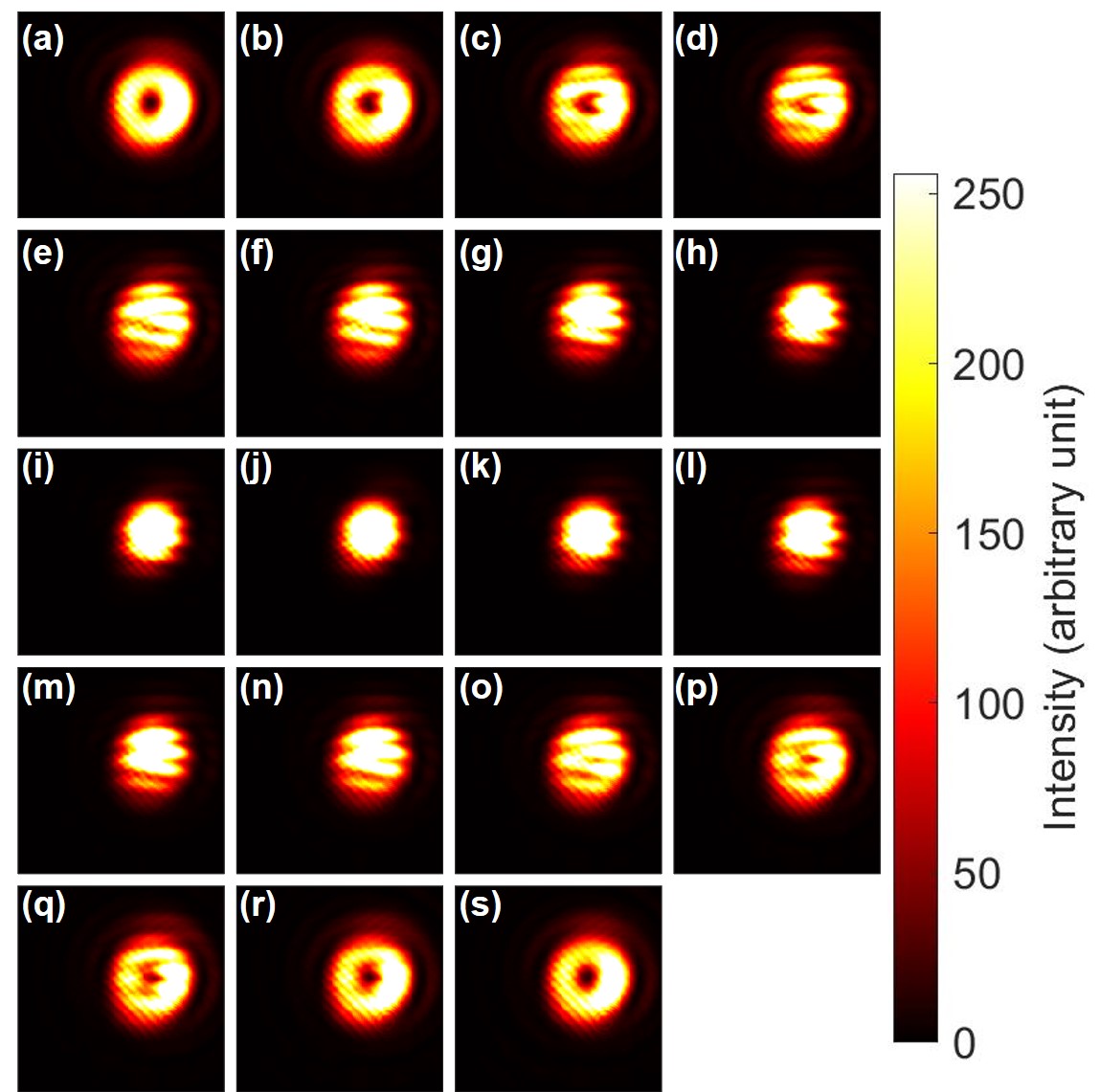}
\caption{
Interference fringes between the left vortexed state and no-vortex state with the intensional miss-alignment between rays.
The ratio of the intensities were controlled by rotator operations of Poincar\'e rotator upon rotating the half-wave plate from (a) 0$^{\circ}$ to (s) 90$^{\circ}$ with a step of 5$^{\circ}$.
Far-field images of (a) and (s) a pure left-vortexed beam, and (j) a Gaussian beam are shown, while we see continuous changes of interference fringes between these images.
}
\end{center}
\end{figure}

After confirming the expected operation of Poincar\'e rotator for orbital angular momentum, we have checked interference fringes \cite{Golub07} between a vortexed state and a non-vortex state.
In order to observe the interference, we have removed VL2, while keeping the VL1, and we have intentionally miss-aligned for the ray with the left vortex and the ray without a vortex.
After confirming the interference with equal intensities by the 50:50 splitting at the PBS, we have systematically changed the splitting by our rotator (Fig. 8).

Suppose if we have not used VL1 at all, we should just see continuous lines of interfere patterns for the superposition of 2 rays with the slight miss-alignment, which induces both constructive and destructive interference.
The lines should be looked like a river to show the profile of the continuous phase as the intensity profile.
On the other hand, if we have a topological charge, which induces $2\pi$ change of the phase around it, the number of interface fringes is different for the amount of the topological charge.

In fact, as confirmed in Fig. 8 (c), (d), (p) and (q), we can recognise that 1 of the high intensity region is terminated at the centre of the topological charge, which looked like a folk.
This confirms that we have generated a vortex of photons, whose topological charge, corresponding to the winding number \cite{Allen92,Golub07,Saito20c}, is 1.

\begin{figure}[h]
\begin{center}
\includegraphics[width=8cm]{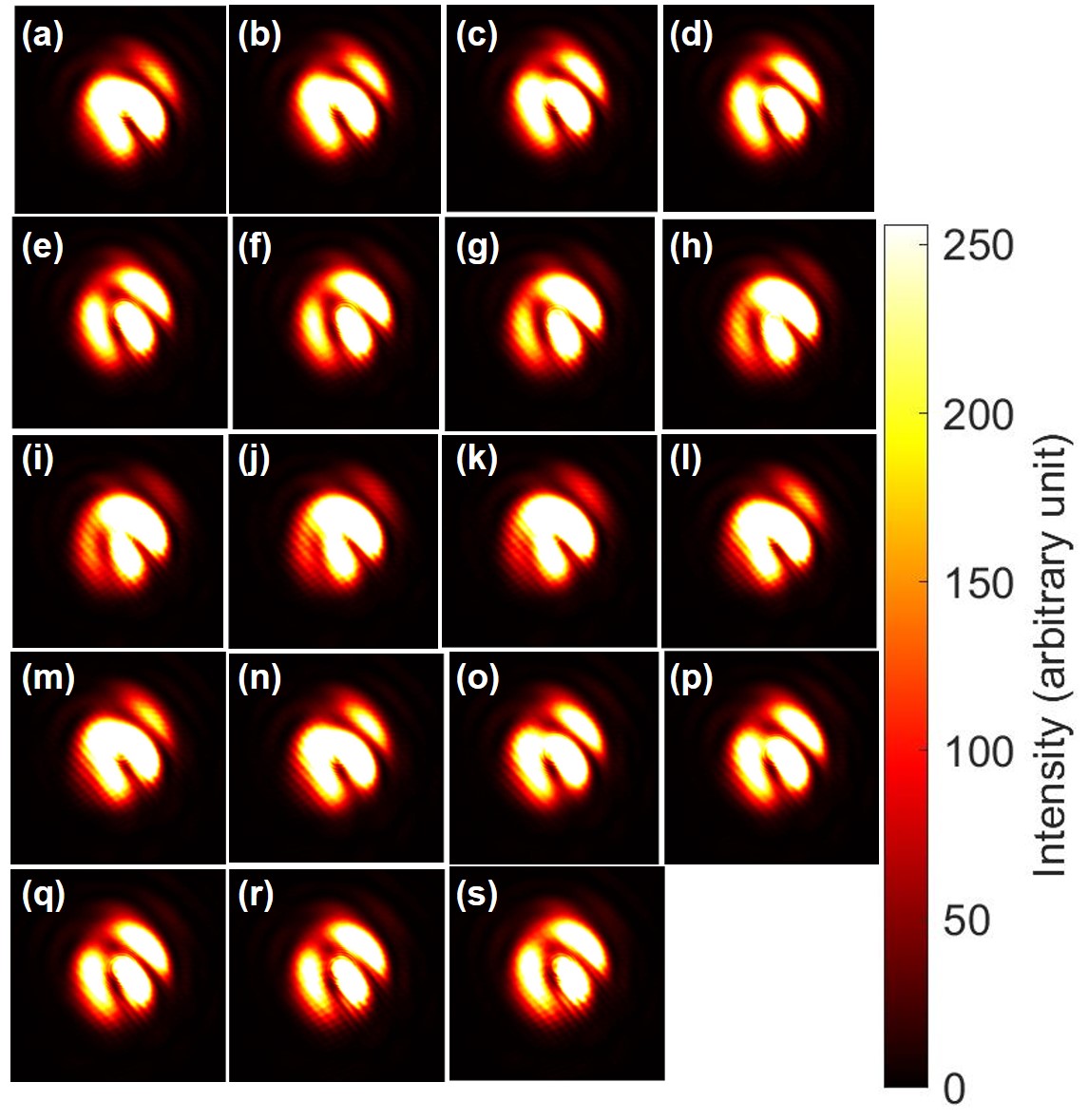}
\caption{
Interference fringes between the left vortexed state and no-vortex state, upon changing the phase by the Poincar\'e rotator.
The phases were controlled by phase-shifter operations of upon rotating the half-wave plate from (a) 0$^{\circ}$ to (s) 90$^{\circ}$ with a step of 5$^{\circ}$.
}
\end{center}
\end{figure}

We have also observed the interference fringes, upon controlling the phase by the Poincar\'e rotator (Fig. 9).
In this experiment, we have fixed the splitting ratio of 50:50 at the PBS, and we have changed the rotation angle of HWP4.
We see that the interference fringes were continuously changed upon the phase-shifts (Fig. 9).
It can also be recognised that the river-like fringe patterns were not affected significantly upon the phase-shifts, since the majority of the interference image is determined by the alignment between 2 rays.
Consequently, the topological charge, which is trapped at the centre of the mode at Fig. 8 (a), could move through the river-like interference fringes to disappear to be a pure Gaussian mode at Fig. 8 (j).
In other words, the river-like interference fringes were pined at the fixed position to determine the boundary conditions for the motion of the topological charge.

\subsection{Topological colour-charged states}

The situation could be changed, if we consider the coupling between a vortexed state and no-vortexed state under properly aligned conditions.
In this case, we could consider coupling among 3 orthogonal states, a left-vortexed state ($|{\rm L}\rangle$), a right-vortexed state ($|{\rm R}\rangle$), and a standard Gaussian beam with no vortex ($|{\rm O}\rangle$), while we assume the polarisation state is the same.
This corresponds to consider SU(3) states, which would be equivalent to assign topological colour charge to the orthogonal modes \cite{Saito23j}.
We have theoretically analysed the SU(3) states and found that 2 sets of SU(2) rotations can cover the whole Hilbert space, since the $\mathfrak{su}(3)$ algebra is rank of 2 \cite{Pfeifer03,Georgi99,Saito23j}.
We have already confirmed 1 set of such SU(2) rotations for coupling between $|{\rm L}\rangle$ and $|{\rm R}\rangle$.
As a next step, we show SU(2) rotations for coupling between $|{\rm L}\rangle$ and $|{\rm O}\rangle$.

This corresponds to make a superposition state among states with different topological charge, which must be orthogonal each other.
The topological charge must be robustly protected against the deformation of the mode, since it is topological, while a standard Gaussian mode does not contain the topological charge, such that the topological charge must be disappeared upon changing the relative amplitudes.
The question is how this change would be realised.
The mode shape with a topological charge is similar to the shape of a doughnut, while the mode shape of the Gaussian mode is similar to a ball.
Our question is how a doughnut could be changed to a ball, or {\it vice versa}, which should not be achieved by topologically continuous deformation.

\begin{figure}[h]
\begin{center}
\includegraphics[width=8cm]{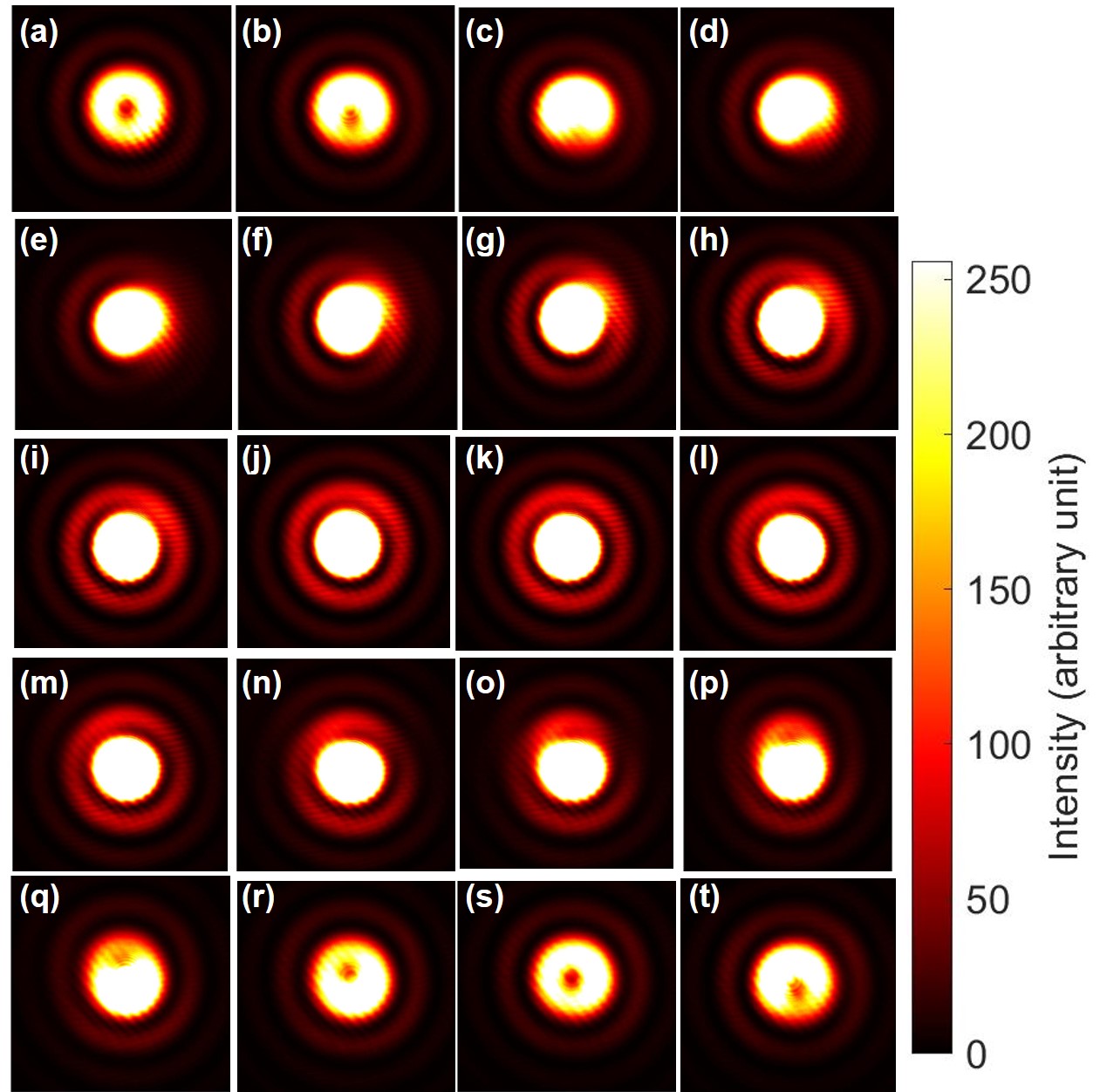}
\caption{
Superposition states between a left-vortexed state and no vortexed state.
The amplitudes were controlled by rotating half-wave plates with the amount of (a) $0^{\circ}$ to (t) $95^{\circ}$ with a step of $5^{\circ}$.
The polarisation state was fixed to be vertical for all images.
(a) and (s) a purely left-vortexed state, and (j) a purely Gaussian state.
The topological charges escaped from being trapped inside the mode, as seen from (a)-(d) and (q)-(t). 
}
\end{center}
\end{figure}

The experimental answer to this question is shown in Fig. 10.
The topological charge moved to escape from being trapped inside the core of the mode.
In fact, the standard mathematical constraint on avoiding to cut the doughnut-like shape is not applicable to our experiments, and the topological charge could cut the edges of the doughnut to leave from the mode (Fig. 10 (c) and (q)).
In fact, upon changing the relative amplitudes of $|{\rm L}\rangle$ and $|{\rm O}\rangle$, some region of the mode will be enhanced due to the constructive interference, while other region will be reduced by the destructive interference to cut the edge, where the topological charge can escape.
The cut-line position is determined by the relative phase, which is not pinned by the interference fridges like those shown in Figs. 8 and 9.
In the properly aligned set-up, there is no interference fringe any more, such that the position for the topological charge to escape must be controlled by the phase-shifter, due to the rotational symmetry of the mode.

\begin{figure}[h]
\begin{center}
\includegraphics[width=8cm]{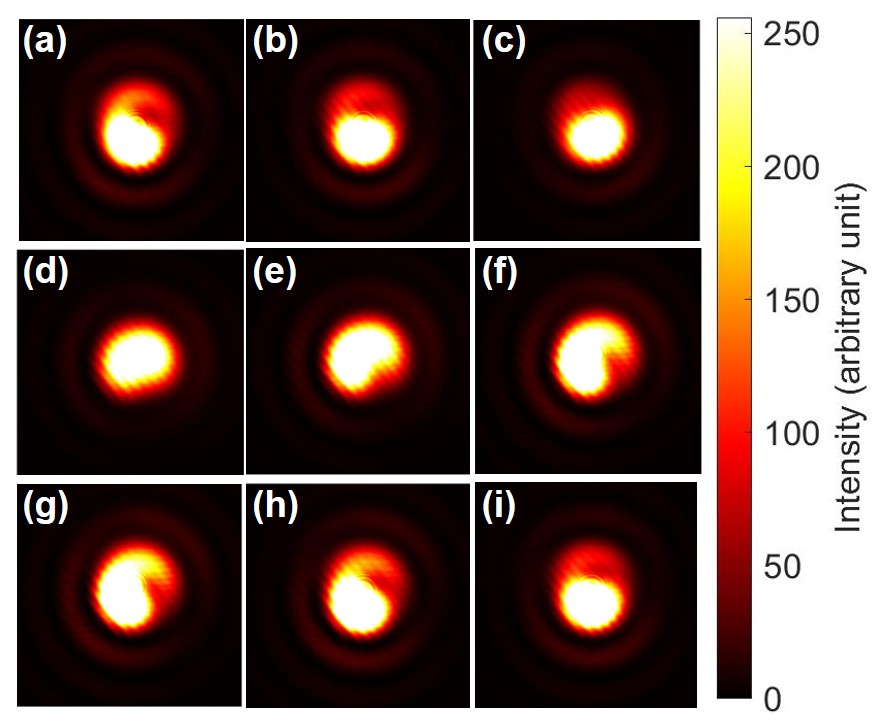}
\caption{
Superposition states between a left-vortexed state and no vortexed state, controlled by the phase-shifter.
The phase was controlled by rotating half-wave plates with the amount of (a) $0^{\circ}$ to (i) $180^{\circ}$ with a step of $22.5^{\circ}$.
The amplitudes were fixed to be 50:50, separated at the polarisation beam splitter, while the polarisation was set to be vertical for both rays.
The topological charge has just escaped at the edge, whose position was rotated upon the phase-shifts.
}
\end{center}
\end{figure}

In order to see the change upon the phase-shifts, we have fixed the splitting at the PBS to be 50:50, and controlled the phase-shifts by rotating HWP4, as shown in Fig. 11.
We see that the topological charge has just escaped from the edge, and the position seems to be rotated upon changing the phase.
In particular, it should be recognised, by comparing Fig. 11 (a) and (i), that the $180^{\circ}$ rotation is required to rotate the bright region at the bottom of the superposition state.
This is again coming from twofold coverage of SU(2) to SO(3), and it is required to rotate twice on the Poincar\'e sphere to bring the phase back to the original state.
This is the same as the rotation of the polarisation state, which makes the horizontally polarised electric field $({\mathcal E}_x,{\mathcal E}_y)=(E_0,0)$ to $(-E_0,0)$ upon 1 rotation on the Poincar\'e sphere, while rotations of twice is required to be $(E_0,0)$.
It is usually difficult to see the phase difference in the polarisation state, because the spin expectation values are the same for both states with the opposite phase.
For the present case of the superposition state between $|{\rm L}\rangle$ and $|{\rm O}\rangle$, we can easily see the difference of the phase between Fig. 7 (a) and (e), since the interfere converts the phase difference to the intensity profile.

\begin{figure}[h]
\begin{center}
\includegraphics[width=8cm]{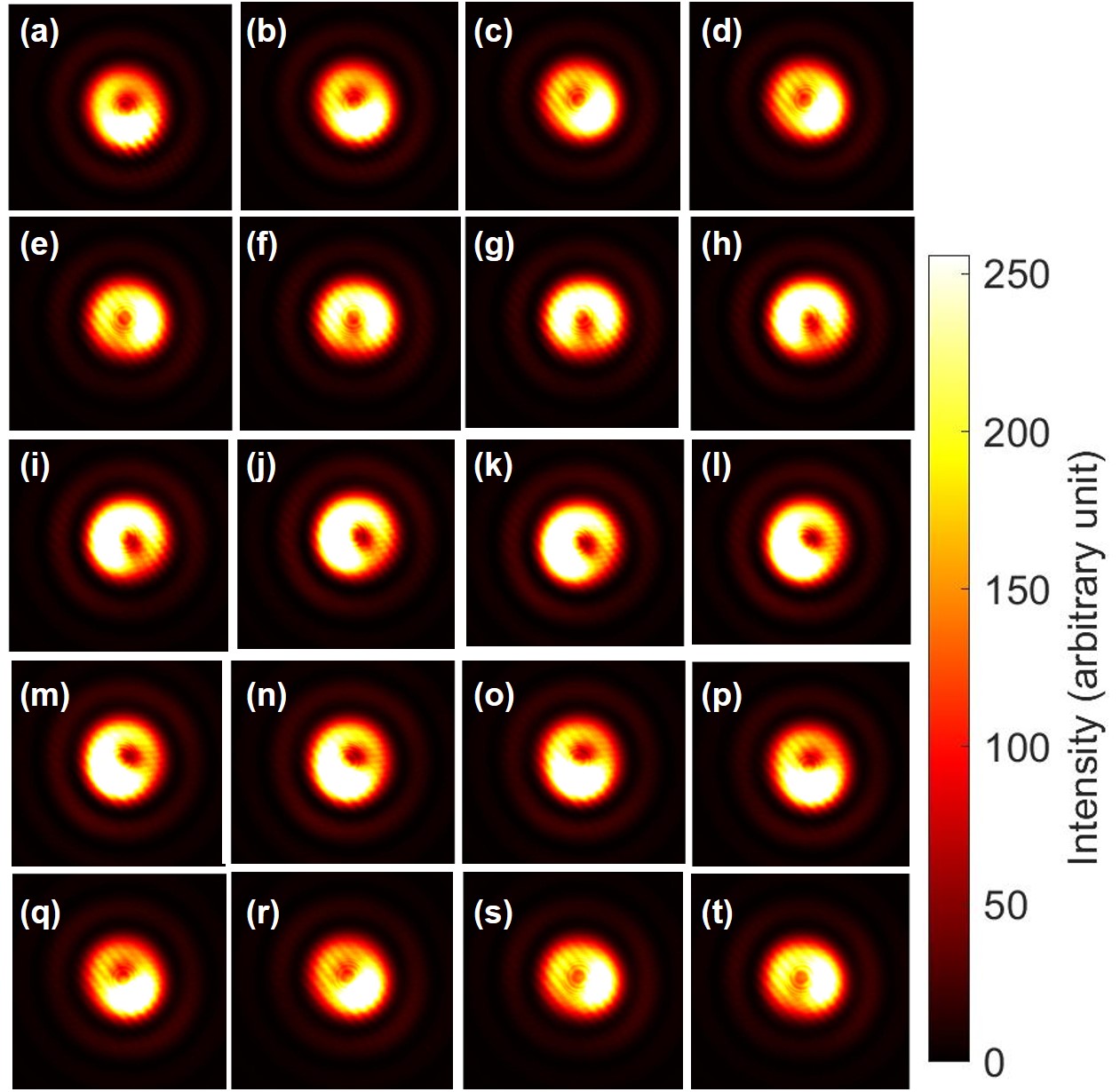}
\caption{
Superposition states between a left-vortexed state and no vortexed state, controlled by the phase-shifter.
The phase was controlled by rotating half-wave plates with the amount of (a) $0^{\circ}$ to (t) $190^{\circ}$ with a step of $10^{\circ}$.
The topological charge is rotating along the anti-clock-wise direction upon increasing the phase-shift, which is expected for the operation of the phase-shifter using Poincar\'e rotator.
}
\end{center}
\end{figure}

On the other hand, it was less clear to see how the topological charge was affected by the phase control from images shown in Fig. 11, we have changed the amplitude by setting HWP1 to align its FA to $10^{\circ}$ from the horizontal direction, and controlled the phase upon rotating HWP4.
As shown in Fig. 11, in this case, we could still see topological charge, located inside the mode, while the position of the topological charge is actually changed upon changing the phase-shift.
The rotation is along the anti-clock-wise direction, which is the same as that for the dipole upon the phase-shift (Fig. 7).

\begin{figure}[h]
\begin{center}
\includegraphics[width=8cm]{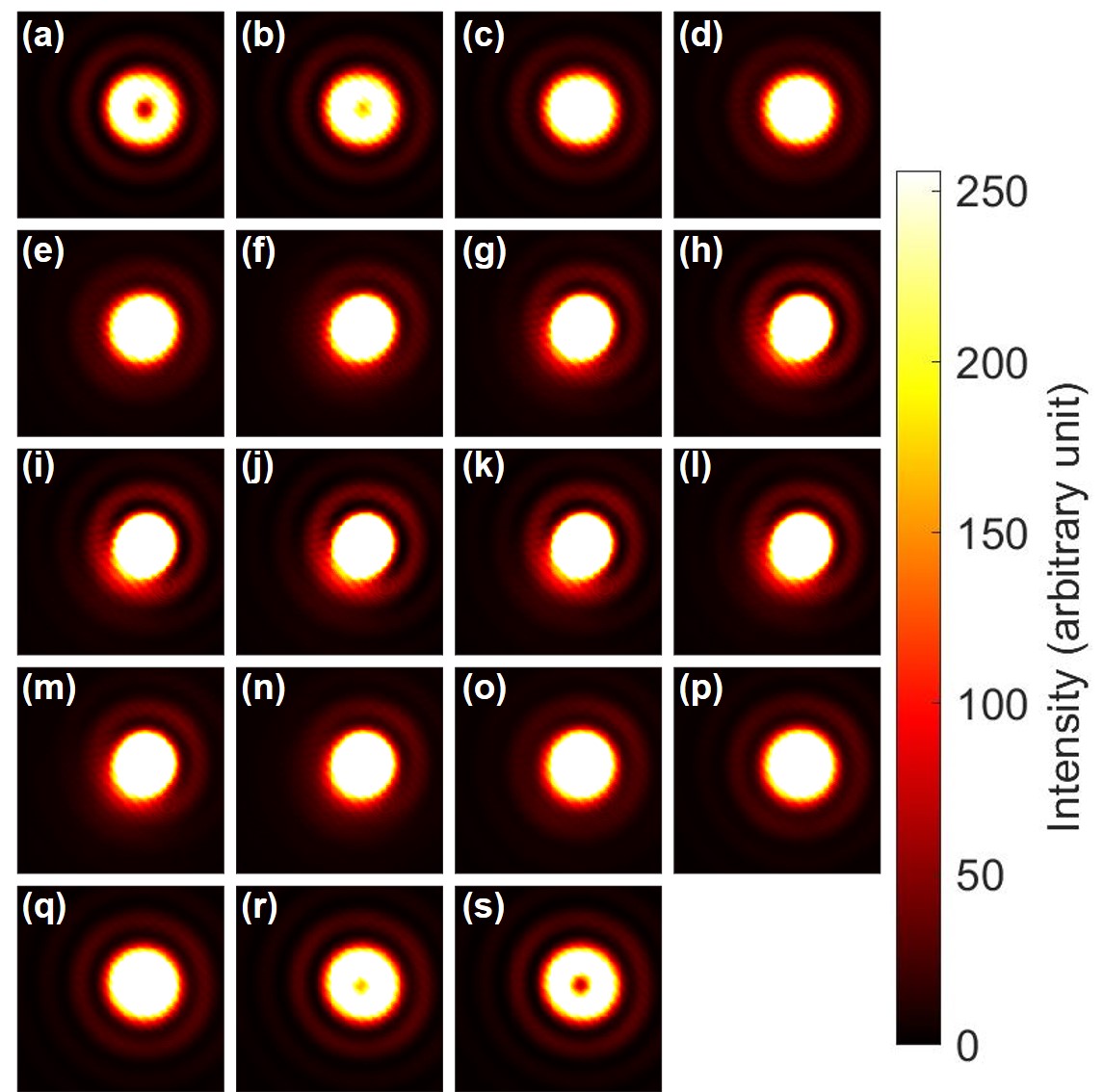}
\caption{
Superposition state between horizontally polarised left-vortexed state and vertically polarised no-vortexed state.
The amplitudes were controlled by rotating half-wave plates with the amount of (a) $0^{\circ}$ to (t) $90^{\circ}$ with a step of $5^{\circ}$.
(a) and (s) a purely left-vortexed state with horizontal polarisation, and (j) a purely Gaussian state with vertical polarisation.
The topological charge was gradually smeared out in (b) and (r), while no obvious interference was seen and the topological charge did not move upon changing the amplitude.
}
\end{center}
\end{figure}

Above demonstrations prove that we can realise an arbitrary superposition state upon mixing 3 orthogonal states of $|{\rm L}\rangle$, $|{\rm R}\rangle$, and $|{\rm O}\rangle$, forming Hilbert space of 3 dimensions, described by SU(3) states \cite{Saito23j}, while we have kept the polarisation state the same vertical state.
The polarisation state is a completely different state, which comes from spin of photons, described by 2 orthogonal states of SU(2) \cite{Saito20a,Saito20b,Saito20c}.
Then, we have also observed the superposition state between horizontally polarised $|{\rm L}\rangle$ and vertically $|{\rm O}\rangle$, as shown in Fig. 13. 
In this case, the interference between orthogonal modes was not observed from the images.
The topological charge stayed at the centre of the mode, and did not move upon changing the amplitudes.

In general, we can prepare 6 orthogonal states for both spin and orbital angular momentum, such that we can explore SU(6) in our experiments.
This will enable us to explore how SU(6) states could be projected into a smaller group of SU(2)$\times$SU(3) for the future.

\subsection{Macroscopic singlet and triplet states}

\begin{figure}[h]
\begin{center}
\includegraphics[width=8cm]{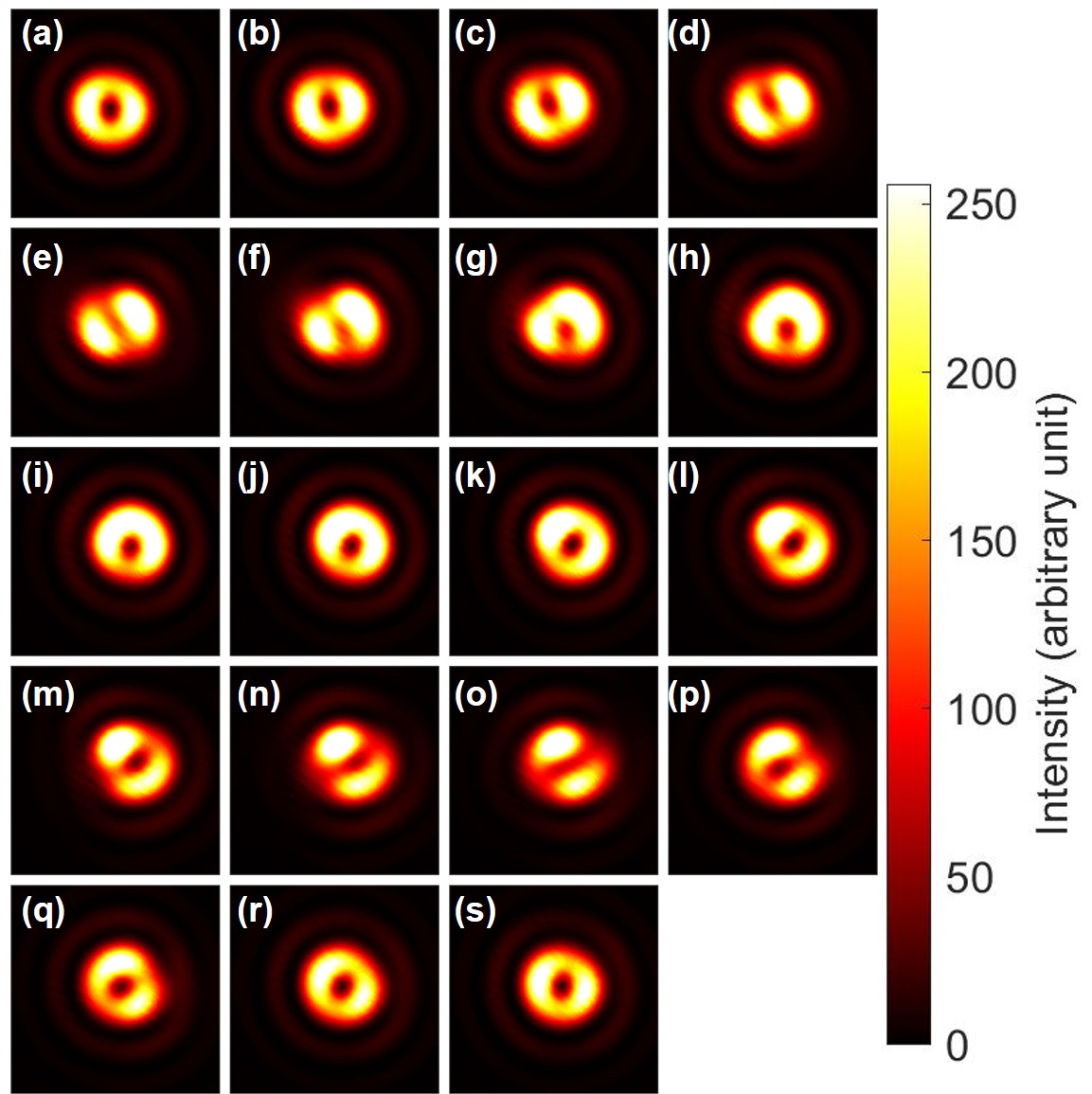}
\caption{
Macroscopic singlet state, realised by the superposition among orthogonal states for both spin and orbital angular momentum.
Polariser was set to rotated to be (a) horizontal at $0^{\circ}$, (j) vertical at $90^{\circ}$, and back to (s) horizontal at $180^{\circ}$ with the step of $10^{\circ}$.
Far-field images are shown for (a) and (s) a pure right vortexed state with the horizontal polarisation, and (j) a pure left vortexed state with the vertical polarisation.
For the diagonal polarisation at (n)-(o), and the anti-diagonal polarisation at (e)-(f), dipoles are pointing along orthogonal directions to the polarisation, respectively, showing total angular momentum of zero.
}
\end{center}
\end{figure}

As for the first step towards considering such a higher order state, we have examined SU(4) states and its projection to SU(2)$\times$SU(2), without using the Gaussian mode without a vortex.
Here, we consider 2 orthogonal polarisation states of horizontal and vertical states as bases for spin, and we also employ left- and right-vortexed states as bases for orbital angular momentum.
Therefore, we have 4 basis states, $|{\rm H}\rangle_{\rm s} |{\rm L}\rangle_{\rm o}$, $|{\rm H}\rangle_{\rm s} |{\rm R}\rangle_{\rm o}$, $|{\rm V}\rangle_{\rm s} |{\rm L}\rangle_{\rm o}$, $|{\rm V}\rangle_{\rm s} |{\rm R}\rangle_{\rm o}$, where ${\rm s}$ and ${\rm o}$ stand for spin and orbital angular momentum, respectively.
In the SU(2) states of polarisation \cite{Stokes51,Poincare92,Jones41,Fano54,Baym69,Sakurai14,Max99,Jackson99,Yariv97,Gil16,Goldstein11,
Hecht17,Pedrotti07,Saito20a}, we can choose the bases to horizontal and vertical states, rather than left and right circularly polarised states.
In the horizontal and vertical bases, and the quantisation axis for the $\hat{\sigma}_3$ operator is attributed to the $S_1$ axis \cite{Saito20a}, such that we will assign spin up and down states as $|{\rm H}\rangle_{\rm s}=|\uparrow \ \rangle_{\rm s}$ and $|{\rm V}\rangle_{\rm s}=|\downarrow \ \rangle_{\rm s}$. 
Similarly, for orbital angular momentum, we assign $|{\rm L}\rangle_{\rm o}=|\uparrow \ \rangle_{\rm o}$ and $|{\rm R}\rangle_{\rm o}=|\downarrow \ \rangle_{\rm o}$.
Consequently, we have 4 orthogonal states
\begin{eqnarray}
&| \ {\rm H} \ \rangle_{\rm s} 
| \ {\rm L} \ \rangle_{\rm o}
=&
|\uparrow \ \rangle_{\rm s}
|\uparrow \ \rangle_{\rm o}
\\
&| \ {\rm H} \ \rangle_{\rm s} 
| \ {\rm R} \ \rangle_{\rm o}
=&
|\uparrow \ \rangle_{\rm s}
|\downarrow \ \rangle_{\rm o}
\\
&| \ {\rm V} \ \rangle_{\rm s} 
| \ {\rm L} \ \rangle_{\rm o}
=&
|\downarrow \ \rangle_{\rm s}
|\uparrow \ \rangle_{\rm o}
\\
&| \ {\rm V} \ \rangle_{\rm s} 
| \ {\rm R} \ \rangle_{\rm o}
=&
|\downarrow \ \rangle_{\rm s}
|\downarrow \ \rangle_{\rm o}
,
\end{eqnarray}
which span the Hilbert space for SU(4) states, since we can realise an arbitrary superposition state with variable phases and amplitudes among preferred combinations of these 4 states.
This corresponds to have 2 quantum bits \cite{Nielsen00,Preskill18}, made of macroscopic quantum states of coherent photons.

In fact, we have already confirmed that our Poincar\'e rotator create an arbitrary SU(2) states for polarisation (Fig. 4), and we have also confirmed that the Poincar\'e rotator together with vortex lenses can generate an arbitrary SU(2) states for orbital angular momentum (Figs. 6 and 7).
However, as far as we are restricted in working to controlling polarisation states for fixed orbital angular momentum or to controlling orbital angular momentum for fixed polarisation states, the Hilbert space is limited for SU(2)$\times$SU(2) states.
In order to realise non-trivial SU(4) states, we need to couple $|\uparrow \ \rangle_{\rm s}|\uparrow \ \rangle_{\rm o}$ and $|\downarrow \ \rangle_{\rm s}
|\downarrow \ \rangle_{\rm o}$, or to couple $|\uparrow \ \rangle_{\rm s} |\downarrow \ \rangle_{\rm o}$ and $|\downarrow \ \rangle_{\rm s} |\uparrow \ \rangle_{\rm o}$.
The most famous states for such combinations would be the singlet state
\begin{eqnarray}
| \ {\rm Singlet} \ \rangle 
&=&
\frac{1}{\sqrt{2}}
\left(
|\uparrow \ \rangle_{\rm s}
|\downarrow \ \rangle_{\rm o}
-
|\downarrow \ \rangle_{\rm s}
|\uparrow \ \rangle_{\rm o}
\right),
\end{eqnarray}
and the triplet states
\begin{eqnarray}
| \ {\rm Triplet} \ \rangle 
&=&
\frac{1}{\sqrt{2}}
\left(
|\uparrow \ \rangle_{\rm s}
|\downarrow \ \rangle_{\rm o}
+
|\downarrow \ \rangle_{\rm s}
|\uparrow \ \rangle_{\rm o}
\right),
\end{eqnarray}
together with $|\uparrow \ \rangle_{\rm s}|\uparrow \ \rangle_{\rm o}$ and $|\downarrow \ \rangle_{\rm s} |\downarrow \ \rangle_{\rm o}$.
In other words, we will realise the coupling of spin and orbital angular momentum, and control the Clebsch-Gordan coefficients \cite{Baym69,Sakurai14}.

In order to realise the singlet and the triplet states for our experiments, we need to use both VL1 and VL2 to generate left- and right-vortexed state, respectively. 
We have aligned HWP5 to align its FA to the $45^{\circ}$ direction from the horizontal axis for rotating the polarisation from the horizontal state to the vertical state.
Therefore, we have generated $|\downarrow \ \rangle_{\rm s} |\uparrow \ \rangle_{\rm o}$ for the ray going to M1.
Similarly, we have aligned HWP6 to align its FA to the $45^{\circ}$ direction from the horizontal axis for rotating the polarisation from the vertical state to the horizontal state.
This allowed us to generate $|\uparrow \ \rangle_{\rm s} |\downarrow \ \rangle_{\rm o}$ for the ray, going to M2.
We can control the phase between $|\uparrow \ \rangle_{\rm s} |\downarrow \ \rangle_{\rm o}$ and $|\downarrow \ \rangle_{\rm s} |\uparrow \ \rangle_{\rm o}$ by the phase-shifter, comprising of QWP1, HWP3, HWP4, and QWP2, as before, simply by rotating the HWP4.
The relative amplitudes were controlled by the rotator, such that we used HWP1 to set the 50:50 splitting at PBS.
Theoretically, our experimental set-up would allow us to realise both singlet and triplet states by macroscopic coherent photons.

There was a challenge, however, to identify the realisation of the singlet and triplet states, since both left- and right-vortexed states provide indistinguishable images at the CMOS camera.
The singlet and triplet states mean that we can identify the difference in phase, once we identify the polarisation, but again the information on the phase could be disappeared in images.

These issues could be simply overcome by changing bases from horizontal and vertical bases to diagonal and anti-diagonal bases, which are defined by 
\begin{eqnarray}
| \nearrow \ \rangle 
&=&
\frac{1}{\sqrt{2}}
\left(
|\uparrow \ \rangle
+
|\downarrow \ \rangle
\right) \\
| \searrow \ \rangle 
&=&
\frac{1}{\sqrt{2}}
\left(
|\uparrow \ \rangle
-
|\downarrow \ \rangle
\right)
,
\end{eqnarray}
respectively, for both spin and orbital angular momentum.
The singlet state must be remained to be the singlet in diagonal and horizontal bases, as
\begin{eqnarray}
| \ {\rm Singlet} \ \rangle 
&=&
\frac{1}{\sqrt{2}}
\left(
|\searrow \ \rangle_{\rm s}
|\nearrow \ \rangle_{\rm o}
-
|\nearrow \ \rangle_{\rm s}
|\searrow \ \rangle_{\rm o}
\right),
\end{eqnarray}
since the total spin after the coupling must vanish.
On the other hand, the triplet state becomes
\begin{eqnarray}
| \ {\rm Triplet} \ \rangle 
&=&
\frac{1}{\sqrt{2}}
\left(
|\nearrow \ \rangle_{\rm s}
|\nearrow \ \rangle_{\rm o}
-
|\searrow \ \rangle_{\rm s}
|\searrow \ \rangle_{\rm o}
\right).
\end{eqnarray}
These states are distinguishable from images for orbital angular momentum, since the diagonal state and the anti-diagonal state correspond to the images of dipoles, which are pointing diagonal and anti-diagonal directions, respectively.

Thus, in order to confirm the singlet and triplet states, we have inserted the polariser, PL2, in front of the NPBS (Fig. 1 (b)).
PL2 was rotated to change the direction of the projection of the polarisation states, which were monitored by the PM.
Please also note that the polarisation state at the CMOS camera is mirror-reflected by the NPBS, such that the diagonal state at the PM should be considered for the anti-diagonal state at the CMOS camera, and {\it vice versa}.

The experimental images are shown in Fig. 14 for the singlet state.
Here, we have adjusted the phase to be singlet, by setting the polariser to be aligned for the diagonal direction at the CMOS camera, and then, HWP4 was rotated to confirm the dipole is aligned to be anti-diagonal direction, as shown in Fig. 14 (n) and (o).
After setting the phase to be singlet, we have observed the images after projecting the polarisation state by PL2.

Among these images, Fig. 14 (a) and (s) correspond to the horizontally polarised state by the projection from the singlet to $|\uparrow \ \rangle_{\rm s} |\downarrow \ \rangle_{\rm o}$, such that the topological charge must be -1 for the right vortex.
On the other hand, Fig. 14 (j) correspond to the vertically polarised state by the projection to $-|\downarrow \ \rangle_{\rm s} |\uparrow \ \rangle_{\rm o}$, such that the topological charge must be 1 for the left vortex.
Unfortunately, we cannot distinguish the chirality of the vortices nor the phase to determine whether the state is singlet or triplet, as we expected above.

Nevertheless, by rotating PL2, we confirmed expected dipole images.
When we set the PL2 to the diagonal direction, the polarisation state at the CMOS camera must be anti-diagonal due to 1 mirror reflection at the NPBS.
The images corresponding to this situation are shown in Fig. 14 (e) and (f), where dipoles are pointing to the diagonal direction.
This was exactly what expected for the singlet state, since we expected projection to $|\searrow \ \rangle_{\rm s} |\nearrow \ \rangle_{\rm o}$.
Similarly, if we align PL2 to the anti-diagonal direction, the polarisation state at the CMOS camera must be diagonal, and we expected the projection to $-|\nearrow \ \rangle_{\rm s} |\searrow \ \rangle_{\rm o}$, which was confirmed in Fig. 14 (n) and (o), whose dipoles are pointing to the anti-diagonal direction.

\begin{figure}[h]
\begin{center}
\includegraphics[width=8cm]{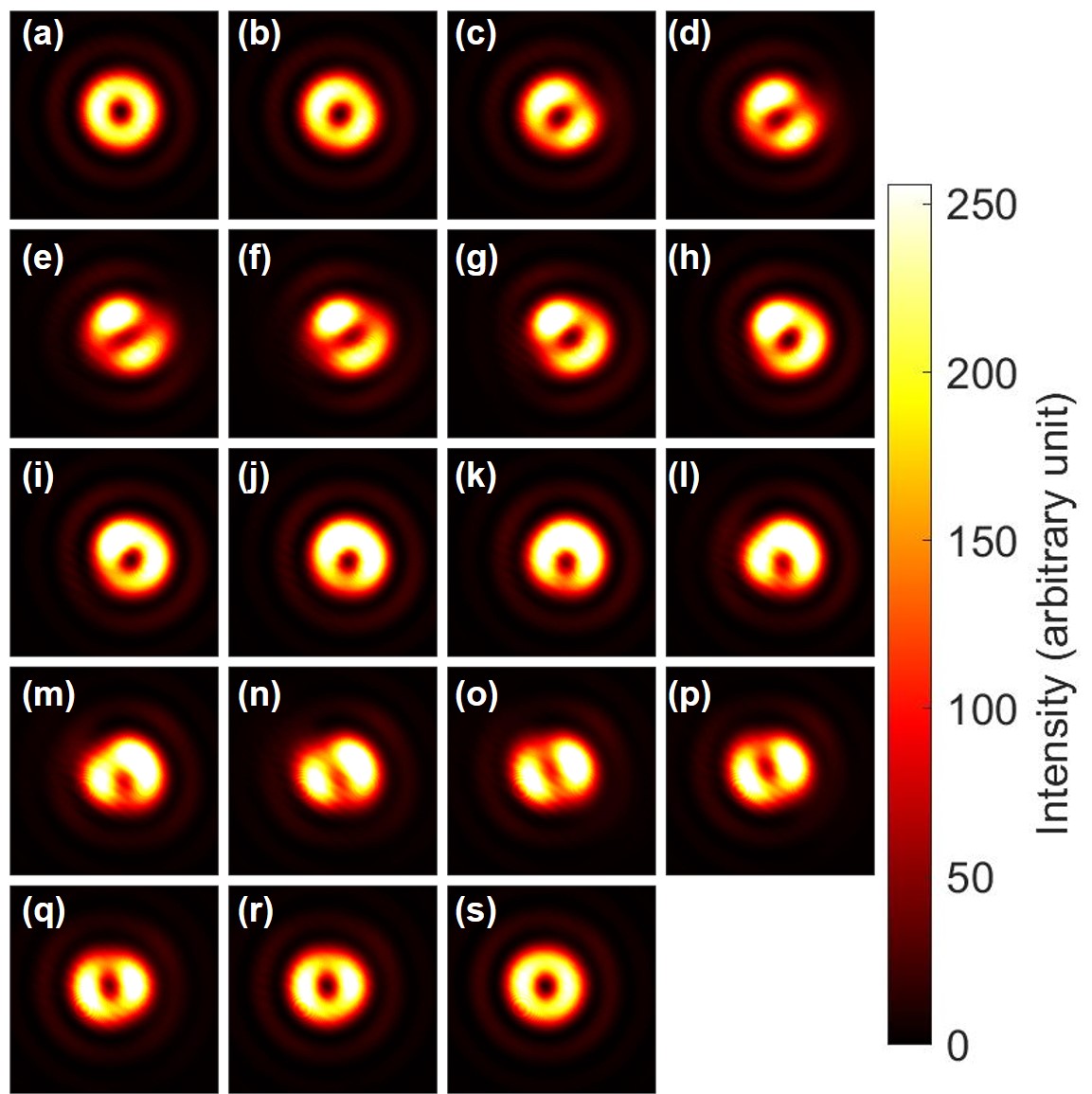}
\caption{
Macroscopic triplet state, realised by the superposition among orthogonal states for both spin and orbital angular momentum.
Polariser was set to rotated to be (a) horizontal at $0^{\circ}$, (j) vertical at $90^{\circ}$, and back to (s) horizontal at $180^{\circ}$ with the step of $10^{\circ}$.
Dipoles are aligned to the same diagonal and anti-diagonal directions with the directions of polarisation.
}
\end{center}
\end{figure}

We have also examined the triplet state.
First, we have set the phase, similar to the singlet state.
After aligning PL2 to the anti-diagonal direction for the polarisation at the CMOS camera, we have rotated HPW4 to align the dipole to the same diagonal direction, seen Fig. 15 (n) and (o).
Then, we have taken images after projecting the polarisation state to the direction defined by PL2, and confirmed that the dipoles are aligned to the same direction with the polarisation, as expected from the simple calculations for SU(4) states, shown above.

We have conducted the physical rotations of PL2, exactly the same way both for the singlet (Fig. 14) and the triplet (Fig. 15) states, however, the rotations in the Poincar\'e sphere for orbital angular momentum are opposite each other.
For the singlet state, the states after the projection were changed from the right-vortexed state,  the diagonal dipole state, the left-vortexed state, and to the anti-diagonal dipole state.
Contrary, for the triplet state, the states after the projection were changed from the right-vortexed state,  the anti-diagonal dipole state, the left-vortexed state, and to the diagonal dipole state.
This difference for the direction of the rotation on the Poinca\'e sphere for orbital angular momentum is simply explained by the SU(4) states, given by the superposition states of two SU(2) states for spin and orbital angular momentum.

\begin{figure}[h]
\begin{center}
\includegraphics[width=8cm]{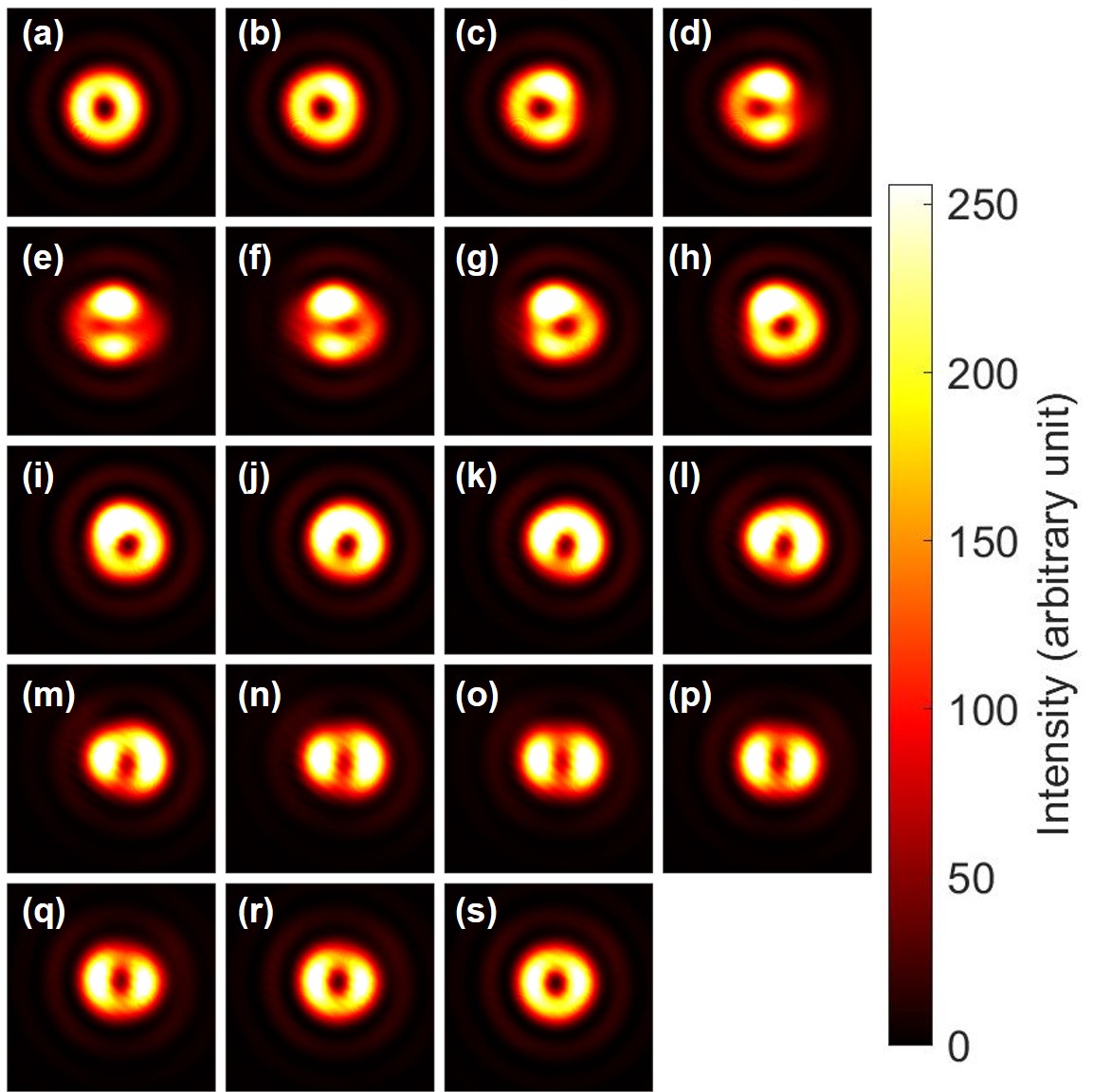}
\caption{
Rotated singlet state of spin and orbital angular momentum.
Polariser was set to rotated to be (a) horizontal at $0^{\circ}$, (j) vertical at $90^{\circ}$, and back to (s) horizontal at $180^{\circ}$ with the step of $10^{\circ}$.
Vertical dipoles are realised under the anti-diagonal polarisation states, and horizontal dipoles are realised under the diagonal polarisation states.
}
\end{center}
\end{figure}

\begin{figure}[h]
\begin{center}
\includegraphics[width=8cm]{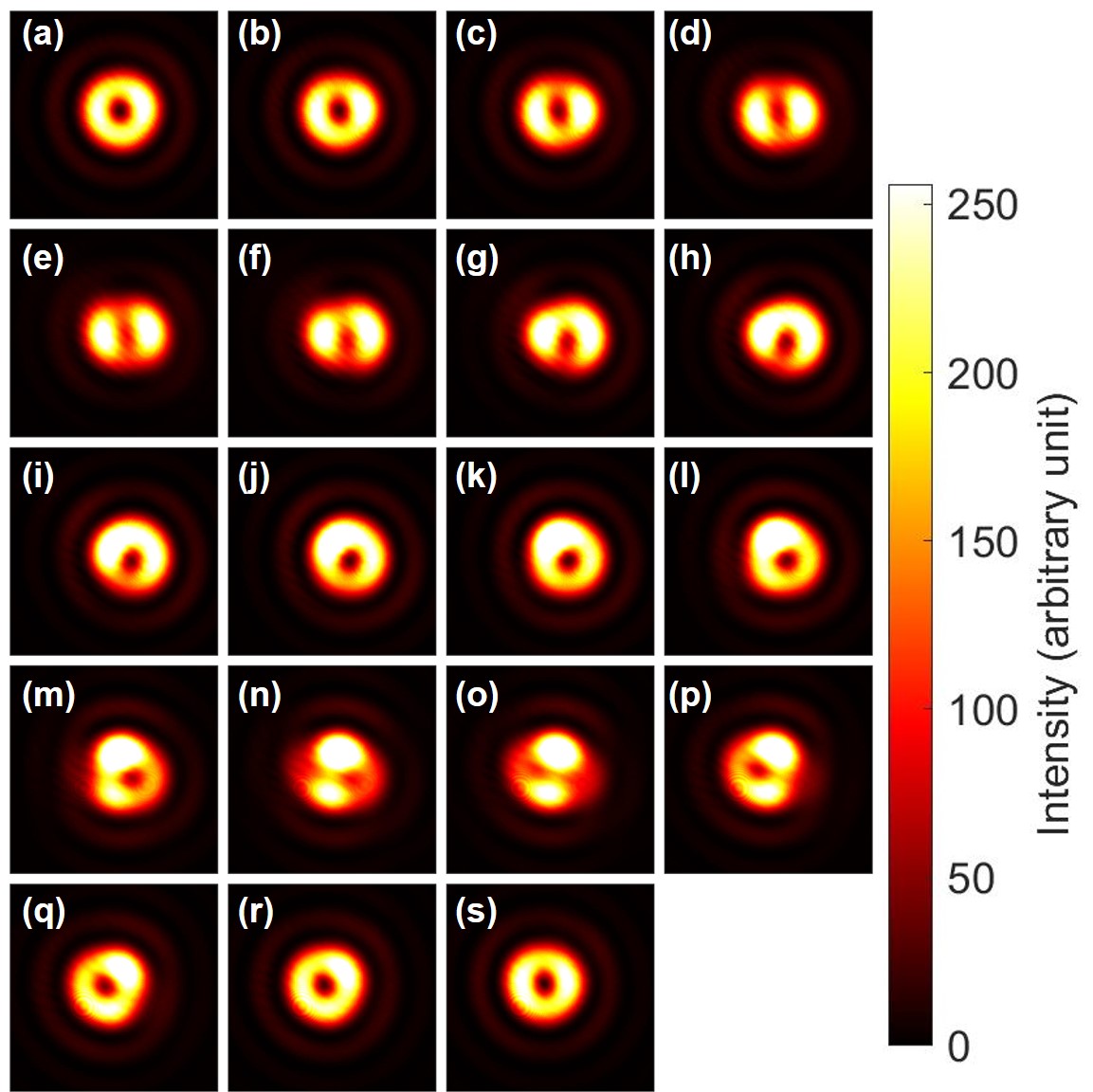}
\caption{
Rotated triplet state of spin and orbital angular momentum.
Polariser was set to rotated to be (a) horizontal at $0^{\circ}$, (j) vertical at $90^{\circ}$, and back to (s) horizontal at $180^{\circ}$ with the step of $10^{\circ}$.
Horizontal dipoles are realised under the anti-diagonal polarisation states, and vertical dipoles are realised under the diagonal polarisation states.
}
\end{center}
\end{figure}

We have also examined the projection from SU(4) states to SU(2)$\times$SU(2) by the polariser for states with horizontal and vertical dipoles, as shown in Figs. 16 and 17, by further phase-shifting the singlet and the triplet states, respectively.
For the experiments, shown in Fig. 16, we have adjusted the phase to align the dipole to the horizontal direction under the diagonally polarisation state (Fig. 16 (n)-(o)).
Then, the rotated singlet state is projected to show the vertical dipole under the anti-diagonally polarisation state (Fig. 16 (e)-(f)).

Similarly, for the experiments in Fig. 17, we have adjusted the phase to align the dipole to the vertical direction under the diagonally polarisation state (Fig. 17 (n)-(o)).
Then, the rotated triplet state is projected to show the horizontal dipole under the anti-diagonally polarisation state (Fig. 17 (e)-(f)).
Therefore, the direction of rotations become opposite between rotated singlet and triplet states in the Poincar\'e sphere for the orbital angular momentum.

All these experiments are consistent with a simple quantum theory of SU(4) states.
After the projection to chose one polarisation state, one of the polarisation state is chosen, and the orbital angular momentum state is also projected by the selection.
The ray after passing through the polariser is described by the SU(2)$\times$SU(2) state, since the ray has certain polarisation state with certain orbital angular momentum state.
The ray in the SU(2)$\times$SU(2) can be still manipulated by wave-plates or other optical components, before observing the states at the CMOS camera.
But, the states of the ray cannot cover the full SU(4) states, spanned by the 4 orthogonal states, since they are not described by the superposition state of orthogonal basis states of SU(4), any more.
In order to recover the full symmetry of SU(4), we need the set up similar to the whole experiment, shown in Fig. 1 (b), to allow the splitting of the ray to recombine after rotational operations for spin and orbital angular momentum.
Mathematically, operators in SU(2)$\times$SU(2) are block diagonalised to conduct separate rotations for spin and orbital angular momentum, individually, while the SU(4) operations are necessary to realise singlet and triplet states.
Our experiments may serve as a platform to explore the projection from SU(4) to U(2)$\times$SU(2).
If we include the non-vortexed state, we can realise a topological colour-charged states to realise SU(3) states for fixed polarisation. 
If we include 2 orthogonal polarisation states, we can prepare 6 orthogonal states to realise SU(6). 
In this case, we can explore the projection from SU(6) to SU(2)$\times$SU(3).
SU(6) is employed for a unified theory in elementary particle physics, while our experiments would be completely different from the mechanism related to the spontaneous symmetry breaking of the vacuum due to a phase-transition \cite{Higgs64,Goldstone62,Nambu59,Abrikosov75,Fetter03,Wen04,Nagaosa99,Altland10,
Gell-Mann61,Gell-Mann64,Ne'eman61,Pfeifer03,Sakurai67,Georgi99,Weinberg05}.
Nevertheless, it might provide some insights to understand the symmetry breaking and associated states, since our projection scheme is experimentally very easy and straightforward to observe the projected states as images.

\section{Conclusion}

We have experimentally demonstrated that spin and orbital angular momentum of coherent photons are described by a representation theory of Lie algebra and Lie group \cite{Stokes51,Poincare92,Stubhaug02,Fulton04,Hall03,Pfeifer03,Dirac30,Jones41,Fano54,Baym69,Sakurai14,Max99,Jackson99,Yariv97,Gil16,Goldstein11,
Hecht17,Pedrotti07,Saito20a,Saito20b,Saito20c,Saito20d,Saito20e,Saito21f,Saito22g,Saito22i,Saito23j}. 
The expectation values of spin orbital angular momentum for the coherent photons are described by Stokes parameters, and we have experimentally developed a simple technique to control the polarisation states by the combination of quarter- and half-wave plates over the full Poincar\'e sphere.
We have also demonstrated that we can also transfer the spin wavefunction to the orbital wavefunction by vortex lenses, such that the full control of the expectation value of photonic orbital angular momentum is also possible.
By extending this technique, we have also demonstrated to realise a superposition state between a vortexed state and a Gaussian mode, and we have observed the continuous motion of the topological charge upon changing the amplitude and the phase of the wavefunction.
This allows us to prepare SU(3) states, which can be assigned to topological colour-charge for coherent photons, which are characterised by the Gell-Mann parameters, which are expectation values of $\mathfrak{su}(3)$ generators of rotation and could be represented on the Gell-Mann hypersphere in SO(8) \cite{Saito23j}.
By combining spin and orbital angular momentum, we could construct both singlet and triplet states, which were confirmed by the projection of the spin state by the polariser and the observed images were consistent with theoretical expectations.
Our results show that we can apply a standard quantum mechanics for spin and orbital degree of freedom for coherent photons.
If we apply our technique to quantum technologies, our results show that we can construct at least 2 qubits by using coherent photons, emitted from a laser diode.

\section*{Acknowledgements}
This work is supported by JSPS KAKENHI Grant Number JP 18K19958.
The author would like to express sincere thanks to Prof I. Tomita for continuous discussions and encouragements.

\bibliography{SO8}

\end{document}